\documentclass[prd,superscriptaddress,nofootinbib,twocolumn]{revtex4}
\usepackage{amssymb,amsmath,graphicx,amscd,epsfig}
\usepackage[colorlinks=true, pdfstartview=FitV, linkcolor=blue, citecolor=red, urlcolor=magenta, breaklinks=true]{hyperref}

\usepackage[T1,T5]{fontenc}
\usepackage{graphicx}
\usepackage[all]{xy}
\usepackage{amsmath}
\usepackage{amssymb}
\usepackage{color}
\usepackage{subeqnarray}
\usepackage{subfigure}
\usepackage{epstopdf}

\newcommand{\bb}{\bibitem}
\newcommand{\bes}{\begin{subequations}}
\newcommand{\ees}{\end{subequations}}
\def\ben{\begin{eqnarray}}
\def\een{\end{eqnarray}}
\newcommand{\bens}{\begin{subeqnarray}}
\newcommand{\eens}{\end{subeqnarray}}

\def\be{\begin{equation}}
\def\ee{\end{equation}}
\def\e{\text{e}}

\def\tanh{\text{tanh}}
\def\sech{\text{sech}}

\def\cos{\text{cos}}

\begin{document}

\title{Scalar fields, localized structures and the Starobinsky model}

\author{D. Bazeia}\email{dbazeia@gmail.com}
\affiliation{Departamento de Física, Universidade Federal da Paraíba, 58051-970 João Pessoa, PB, Brazil}

\author{Elisama E. M. Lima} \email{elisama.lima@ifba.edu.br}
\affiliation{Instituto Federal de Ciência e Tecnologia da Bahia, 47808-006, Barreiras, BA, Brazil}

\begin{abstract} 
This work deals with the presence of localized static structures in the real line, described by relativistic real scalar fields in two spacetime dimensions. We consider models featuring both standard and modified kinematics, where we employ two intriguing potentials supporting defect solutions. The first potential can transform kink into compacton in the standard framework,  while the second one is based on the inflationary Starobinsky model. Interesting possibilities unseen in previous investigations are described, in particular, for the case related to the Starobinsky potential. The addressed potentials are inserted into a broader framework, and so the extended models are described by a wider set of solutions. This investigation also reveals the presence of the twinlike behaviour for a specific compact configuration, which solves two distinct models.
\end{abstract}

\maketitle

\pretolerance10000

\section{Introduction} \label{intro}

Topological structures hold significant relevance in high energy physics \cite{B1,B2,B3}, as well as in various other branches of nonlinear science; see, e.g., \cite{BB1,BB2,BB3,BB4} and references therein. In the context of high energy physics, these structures are described by field configurations with finite energy, and they can emerge in the form of kinks, vortices and magnetic monopoles, among other possibilities. The equations of motion are satisfied in their respective models and spacetime dimensions. Kinks are perhaps the simplest of such structures, and they are typically formed within the context of real scalar fields in $(1, 1)$ spacetime dimensions.

In this work, we deal with kinklike solutions in two different scenarios. One is based on the standard description of field models, and another is characterized by a generalized kinematics, which includes terms dependent on higher-order power in the first derivative of the fields. Our interest is exploring kinks and compactlike structures \cite{rosenau,compactModels,morecompact} driven by real scalar fields in the context of standard and generalized theories. Unlike traditional kinks, compactlike structures reside within a compact space. A kink presents an energy density that decreases asymptotically to zero, while the energy density of a compact structure is exactly null outside a closed interval.

One aspect of generalized models is that they usually can accommodate compactons due to nonlinearities present there, although in certain situations standard defects can also get a compact behaviour \cite{morecompact}. Here, we adopt two interesting and distinct manners to explore the behaviour of defect solutions and their transformation into compactons. In the first possibility, the potential we choose has the property of transforming kink into compacton in a model with standard kinematics; in the braneworld scenario, it revealed a hybrid profile for a flat brane within five-dimensional spacetime, as detailed in Ref.~\cite{morecompact}. The second case is based on the Starobinsky model, which is a widely celebrated theory in general used to describe dynamics of inflation, see Ref.~\cite{epjp-2021} and references therein for more information on this issue. Although the potentials used to describe these two possibilities have been used in distinct contexts, we show here that they can share some interesting similarities, unseen in the previous investigations.

To simplify our investigation, we introduce an auxiliary function $W = W(\phi)$, which allows us to derive first-order equations in a way similar to the cases explored before in \cite{FOFGD1,FOFGD2,FOFGD3}. These equations greatly facilitate the study and resolution of the equations of motion. The utilization of the deformation procedure, as demonstrated in references \cite{Bazeia:2002xg,almeida,bazeialosano}, has proven to be a valuable route to obtain the examined potentials. See also Refs. \cite{D0,D1,D2,D3} for other uses of the deformation procedure. Initially, we review the main characteristics of the defect solutions with standard kinematics. After, we describe their evolution in response to variations in the dynamics. Then, we introduce a driving parameter that governs the transition from the standard configuration to the generalized model.

In order to deal with all the above issues, we organize the present work as follows: in Sec.~\ref{sec-1} we introduce some general concepts involving both standard and generalized models. In Sec.~\ref{sec-2},  we provide a description of the standard models, emphasizing the principal attributes of the localized solutions. In Sec.~\ref{sec-3}, we extend the standard case by considering two generalized models. We then conclude our investigation with some comments and conclusions in Sec.~\ref{sec-4}.

\section{Generalities} \label{sec-1}

A  single real scalar field model in $(1, 1)$ spacetime dimensions is described by the Lagrange density
\be
\label{lagran1}
{\cal L}=\frac12\partial_\mu\phi\partial^\mu\phi-V(\phi),
\ee
or yet
\be
\label{Lstan}
{\cal L}=X-V(\phi),
\ee
where $X=\frac12\partial_\mu\phi\partial^\mu\phi$. In this work we shall use natural units and also, dimensionless fields and spacetime coordinates. The equation of motion  for static field configurations is
\be
\phi''=V_\phi.
\ee
By defining a function $W(\phi)$ such as $V(\phi)=\frac12 W_{\phi}^2$, the equation can be reduced to first-order
\be
\phi'=W_{\phi}. 
\ee

Let us now consider a more general model such that
\be
\label{lagran}
{\cal L}\left(\phi,X\right)=F(X)-V(\phi),
\ee
where $F(X)=X$  is used to restore the standard Lagrange density.  As detailed in \cite{FOFGD1,FOFGD2}, for static field configurations $\phi(x)$, 
the equation of motion and the stressless condition assume the general forms
\be
(F_X\phi')'=V_\phi
\ee
and
\be
\label{Vgeneral}
V=F-2F_X X.
\ee 

The first-order framework  can be defined as 
\be
F_{X}\phi'=W_{\phi}.
\label{w}
\ee
The energy density assumes the form
\be
\rho(x)=W_\phi \phi'=\frac{d W}{dx},
\label{rhoFx}
\ee
in such a way that the energy can be written in terms of the variation of $W$ at the asymptotic limits
\be
E=W(\phi(x\rightarrow \infty))-W(\phi(x\rightarrow -\infty)). \nonumber \\
\ee

The linear stability is analyzed by assuming small perturbations around the static solution, $\phi(x,t)=\phi(x)+\eta_n(x)\cos(\omega_n t)$. An equation similar to Schrödinger's equation is obtained using the stability potential
\be
\label{spz}
U(z)=\frac{(\sqrt{F_{X}A})_{zz}}{\sqrt{F_{X}A}}+\frac{V_{\phi\phi}}{F_X},
\ee
where the variable $x$ has been changed according to $dx = Adz$, and $A^2=(2XF_{XX}+F_X)/F_X$, as previously described in \cite{FOFGD1,FOFGD2}. 

The first situation considered in this work is the standard Lagrangian \eqref{Lstan}. The stability potential, $U(x)$, is expressed as $V_{\phi\phi}$; using the first-order formalism, it can also be written in terms of the function $W$ 
\be
V_{\phi\phi}=U(x)=W_{\phi\phi}^2+W_{\phi}W_{\phi\phi\phi},
\ee
which has to be evaluated at the static solution $\phi=\phi(x)$. In this case, the Schrödinger-like equation can be factorized, leading to linearly stable solutions \cite{Bazeia}.

The second case assumes two extensions of the standard dynamics. The first extension considers the kinetic term $F(X)=-X^2$, leading to the first-order equation $\phi'=W_{\phi}^{1/3}$ and the potential $V(\phi)=\frac34W_{\phi}^{4/3}$. The stability potential in this case is given by
\be
U(z)=2W_{\phi}^{-1/3}W_{\phi\phi\phi}, 
\ee
resulting in stable static solutions \cite{FOFGD1,FOFGD2}.

After that, one uses the dynamics $F(X)=X-\alpha X^2$, where $\alpha$ is a positive real parameter. The limit $\alpha$ very small conducts back to the standard situation, whereas $\alpha$ very large guides the dynamics towards the generalized case represented by $F(X)=-X^2$. Our aim here is to establish a global framework encompassing both the standard case and the first generalized scenario, to explore how the kinklike behaviour evolves to a compacton.  This extension incorporates the addressed potentials within a more general context, where the profiles of the localized solutions and energy densities change as $\alpha$ varies.


\section{Standard Kinematics} \label{sec-2}
In the first scenario, we choose the standard field theory with the potential 
\be
V(\chi)=\frac{1}{2}\left(1-\chi^2\right)^2.
\label{pnormal}
\ee
This model presents the kink solution $\chi(x)= \tanh(x)$, with energy density $\rho(x)=\sech^4(x)$ and energy $E=4/3$. The stability potential is the modified P\"oschl-Teller $U(x)= 4-6\,\sech^2(x)$,  which has two bound states $\omega_0=0$ and $\omega_1^2=3$ \cite{Bazeia,outro,teller,PT}.

From this, one can use the deformation procedure introduced in Refs.~\cite{Bazeia:2002xg,almeida,bazeialosano} to find analytical solutions for other scalar field models. The method results in deformed defects starting from a given scalar theory. It enables the construction of new models supporting defect solutions which are analytically expressed in terms of the original one. To achieve the desired deformation, we start from Eq. \eqref{pnormal} whose solution is already known, and then we use a deforming function $f(\phi)$ to connect it to another one, denoted by $V(\phi)$. The potentials are linked through the transformation $\chi \rightarrow f(\phi)$. Consequently, the solution for the deformed one can be obtained by using the inverse function, $\phi(x) = f^{-1}(\chi(x))$. The relation between the original and the deformed potential is
\be\label{SR}
V(\phi)=\frac{V(\chi\rightarrow f(\phi))}{f_{\phi}^2}=\frac{(1- f^2)^2}{2f_{\phi}^2}.
\ee 

\begin{figure}%
\centering
\includegraphics[scale=0.20]{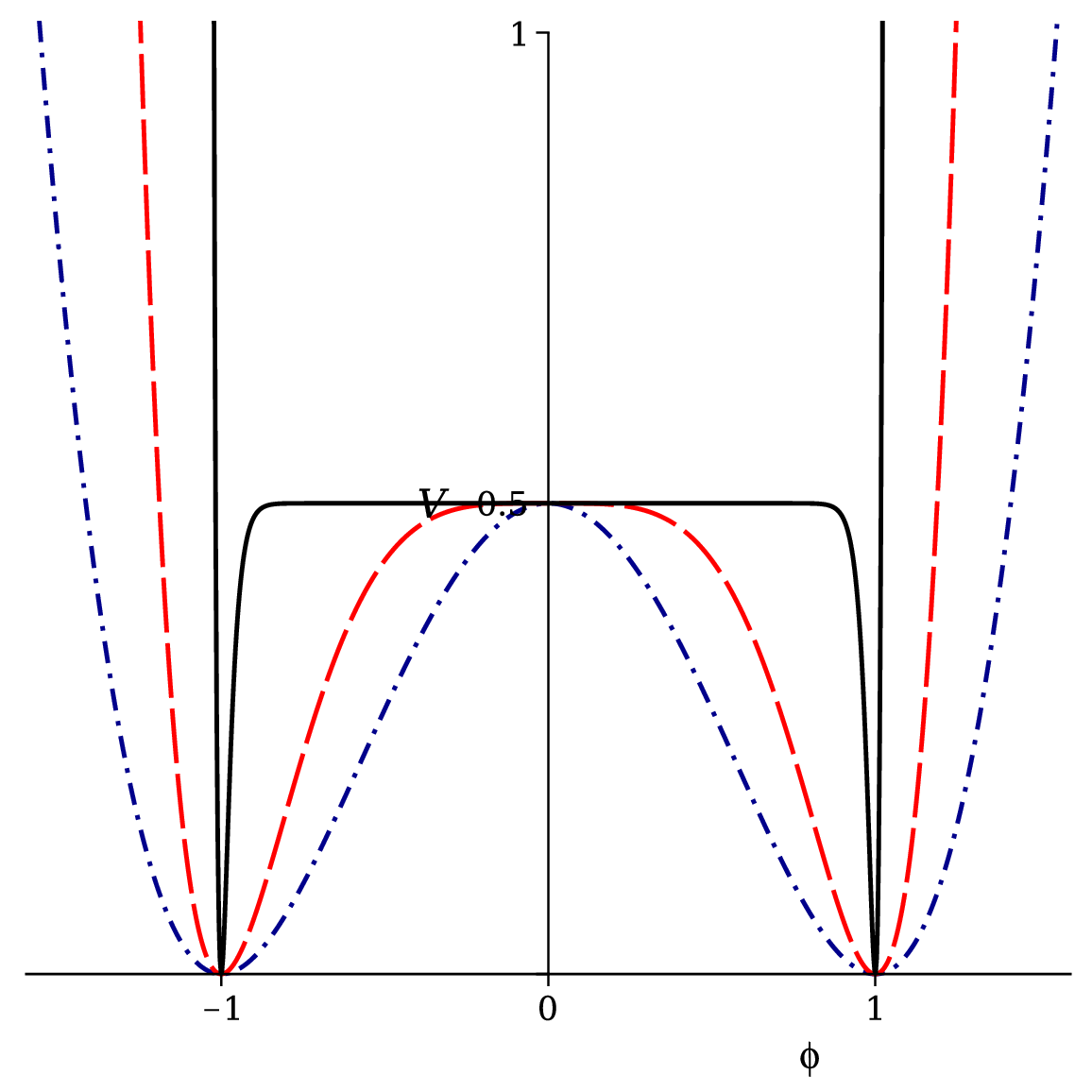}
\includegraphics[scale=0.20]{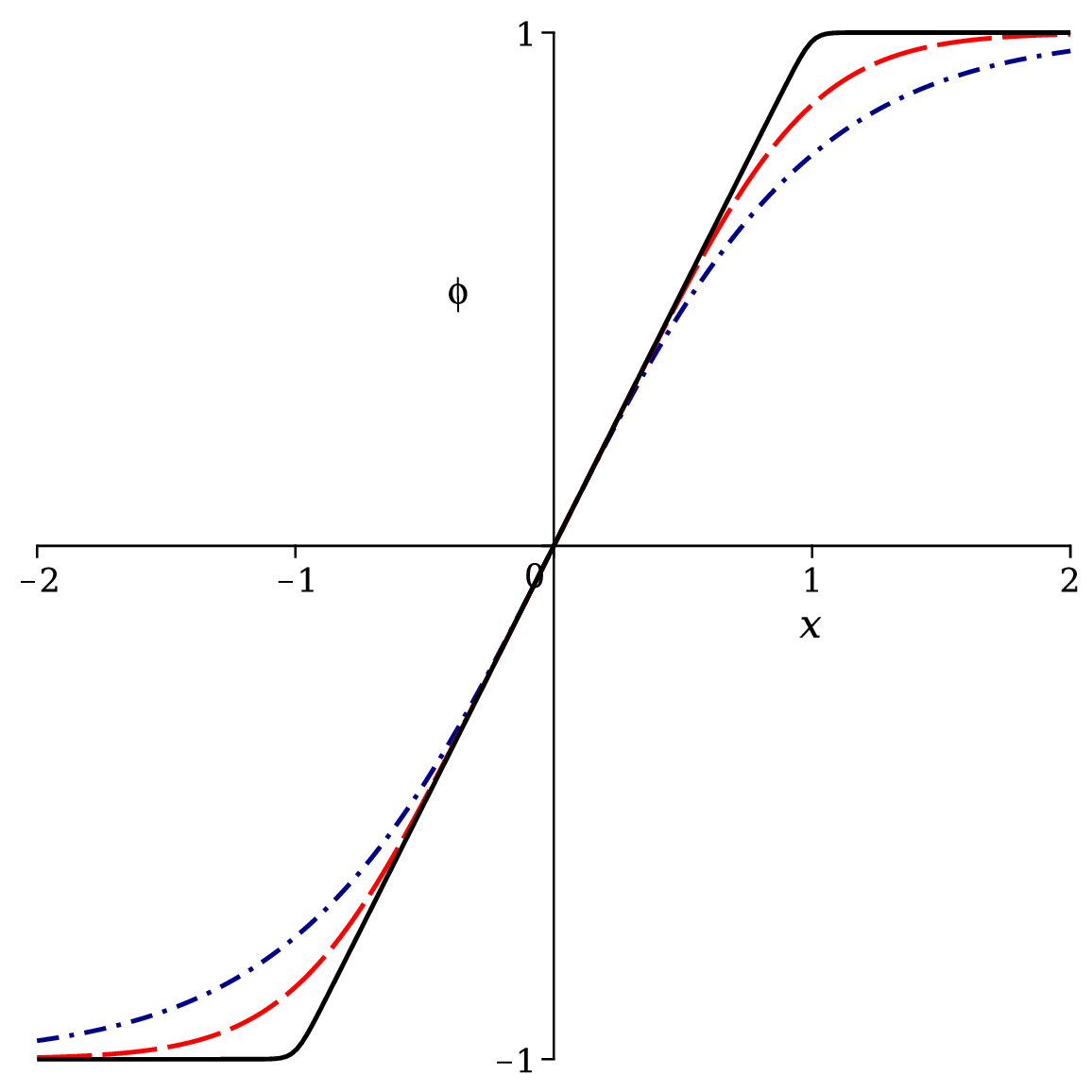}
\includegraphics[scale=0.2]{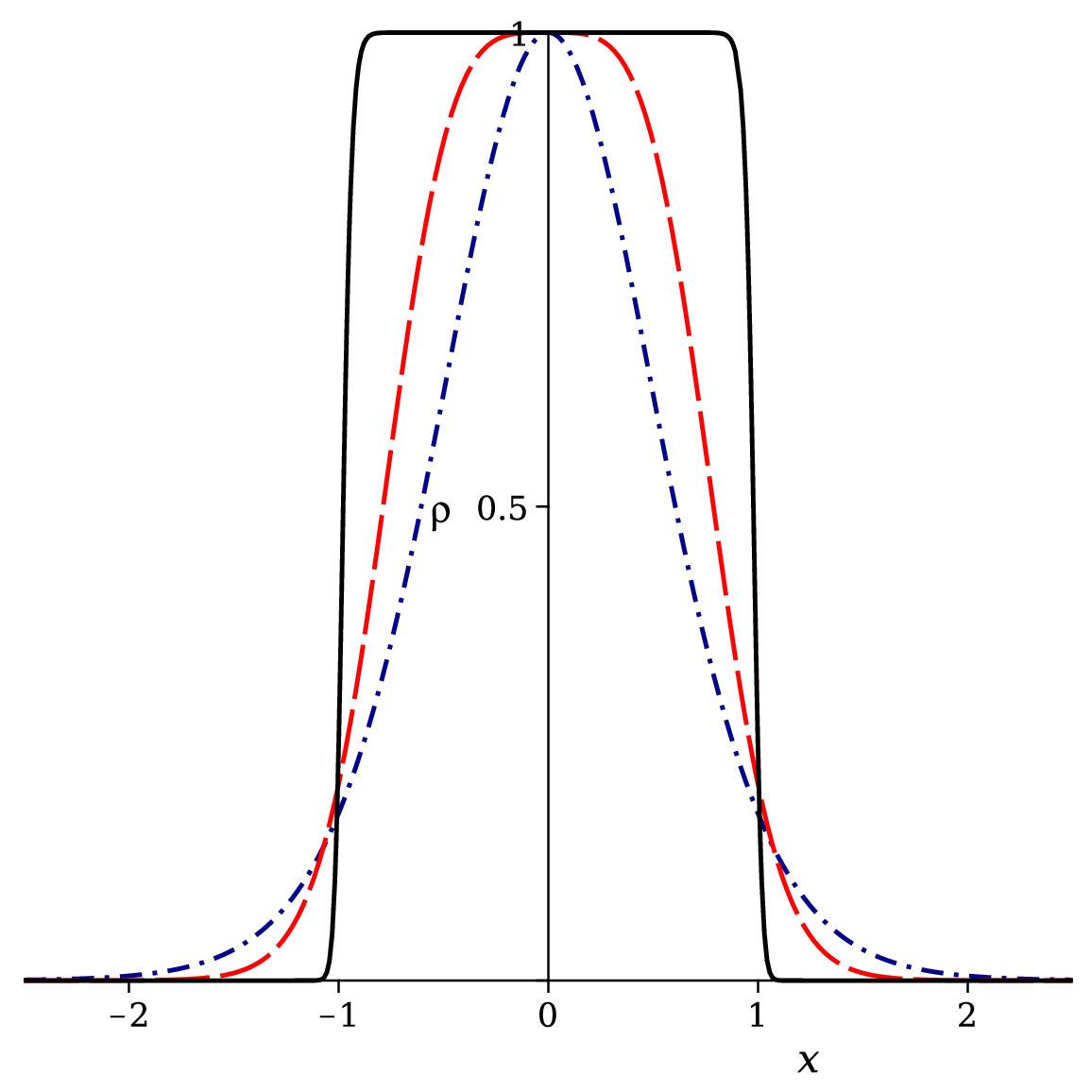}
\includegraphics[scale=0.2]{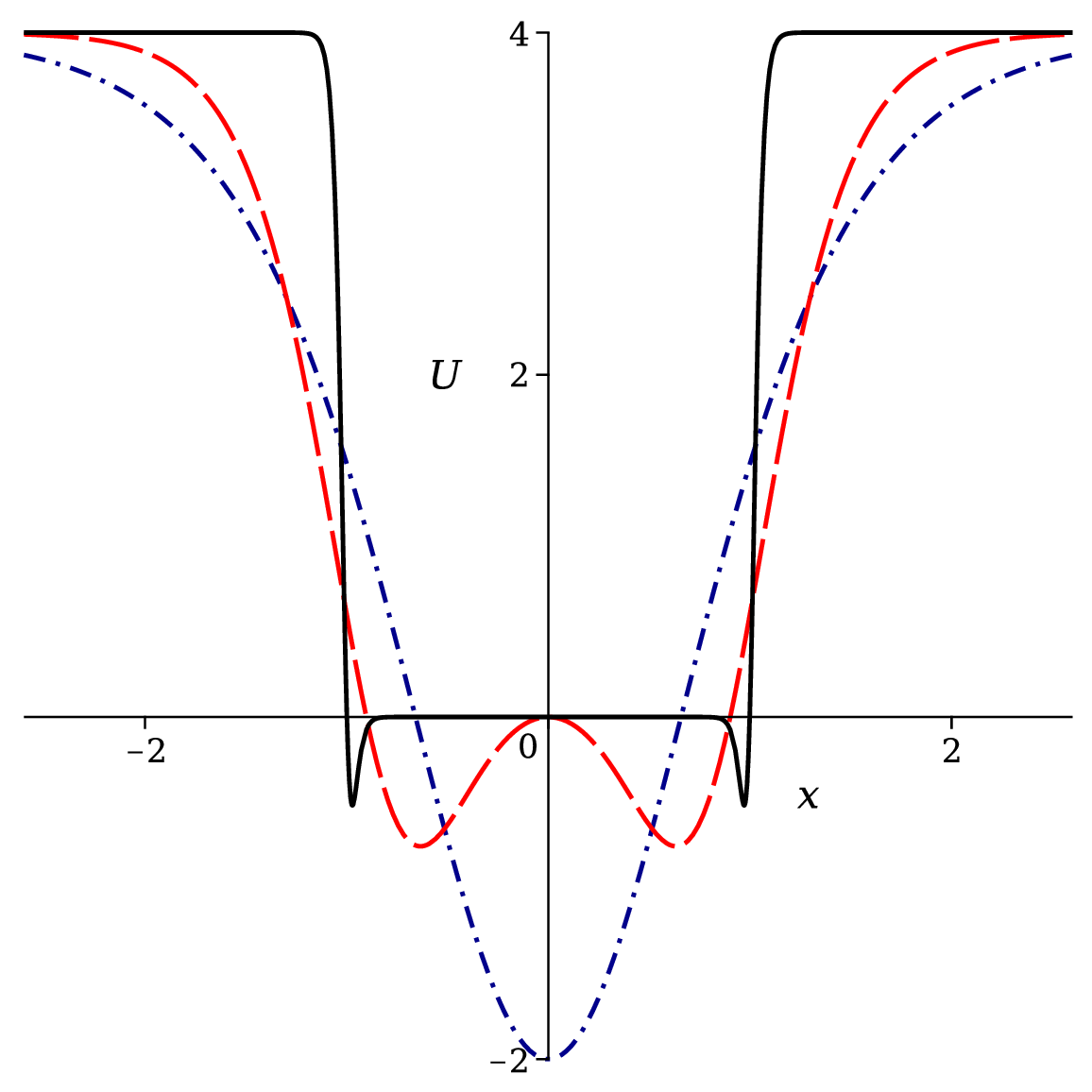}
\caption{The potential $V(\phi)$ given in Eq.~\eqref{Vn} (first panel), the kink solutions $\phi(x)$ (second panel), energy densities $\rho(x)$ (third panel) and stability potentials $U(x)/n^2$ (fourth panel) for $n=1,2,20$, represented by dash-dotted (blue), dashed (red) and solid (black) lines, respectively.}
\label{fig1A}
\end{figure}

The first deformation function suggested is written in terms of the  Lerch transcendent ${\bf \Phi}$ 
\be
f_1(\phi)=\tanh\left(\frac{\phi}{2n}{\bf \Phi} \left(\phi^{2n},1,\frac1{2n}\right)\right)
\ee
where $n$ is an integer positive parameter. The deformed potential becomes
\be
V^{(1)}_n(\phi)=\frac{1}{2}\left(1-\phi^{2n}\right)^2.
\label{Vn}
\ee
In the standard scenario, this model smoothly transforms kinks into compactons as proposed in Ref.~\cite{menezes2014}. For $n = 1$, it recovers the $\phi^4$ model. The minima are at $\phi_{min}=\pm 1$ for any value of $n$, and the mass varies according to $m^2=4n^2$. The first-order equation is
\be
\phi'=1-\phi^{2n}.
\ee
In Ref.~\cite{menezes2014}, the equation of motion is solved numerically, showing that the solution transits smoothly from a kink to a compacton as $n$ increases. Fig.~\ref{fig1A} displays how the properties of \eqref{Vn} change for some values of $n$, including kink solution, energy density, and stability potential. The mass increases with $n$, making the range of the solution decrease more and more until it gets a compact behaviour, and the potential minima are reached at finite values of $x$.

In this work, we show that the first-order formalism and the deformation procedure improve the previous scenario, allowing that we find an analytical expression for the kink solutions by taking the inverse function $\phi(x) = f^{-1}(\tanh(x))$, which gives
\be
\phi\,{\bf \Phi}\left( \phi^{2n},1,\frac1{2n}\right)=2nx.
\label{solVnstandard}
\ee

The energy can be obtained via function $W$, where
\be
W^{(1)}_n(\phi)=\phi-\frac{\phi^{2n+1}}{2n+1}.
\ee
and
\be
E_n^{(1)}=W(1)-W(-1)=\frac{4n}{2n+1}. 
\label{E1}
\ee

Expanding the expression \eqref{solVnstandard} for $n\gg1$, we have
\be
\phi-\frac{\phi}{2n}\ln(1-\phi^{2n})+ {\cal O}\left(\frac{1}{n^2}\right)=x.
\ee 
Thus, in the limit $n \rightarrow \infty$, the solution becomes $\phi(x)=x$ inside the compact interval $x\in[-1,1]$, and the potential minima are reached exactly at $\bar{x}=\pm 1$. The energy density becomes $\rho(x)=1$ inside the compact region and null elsewhere, see the third panel in Fig.~\ref{fig1A}. Moreover, the stability potential is depicted in the fourth panel of Fig.~\ref{fig1A} as $U(x)/n^2$ in order to visualize it more easily, considering the selected values of $n$.

\begin{figure}%
\centering
\includegraphics[scale=0.2]{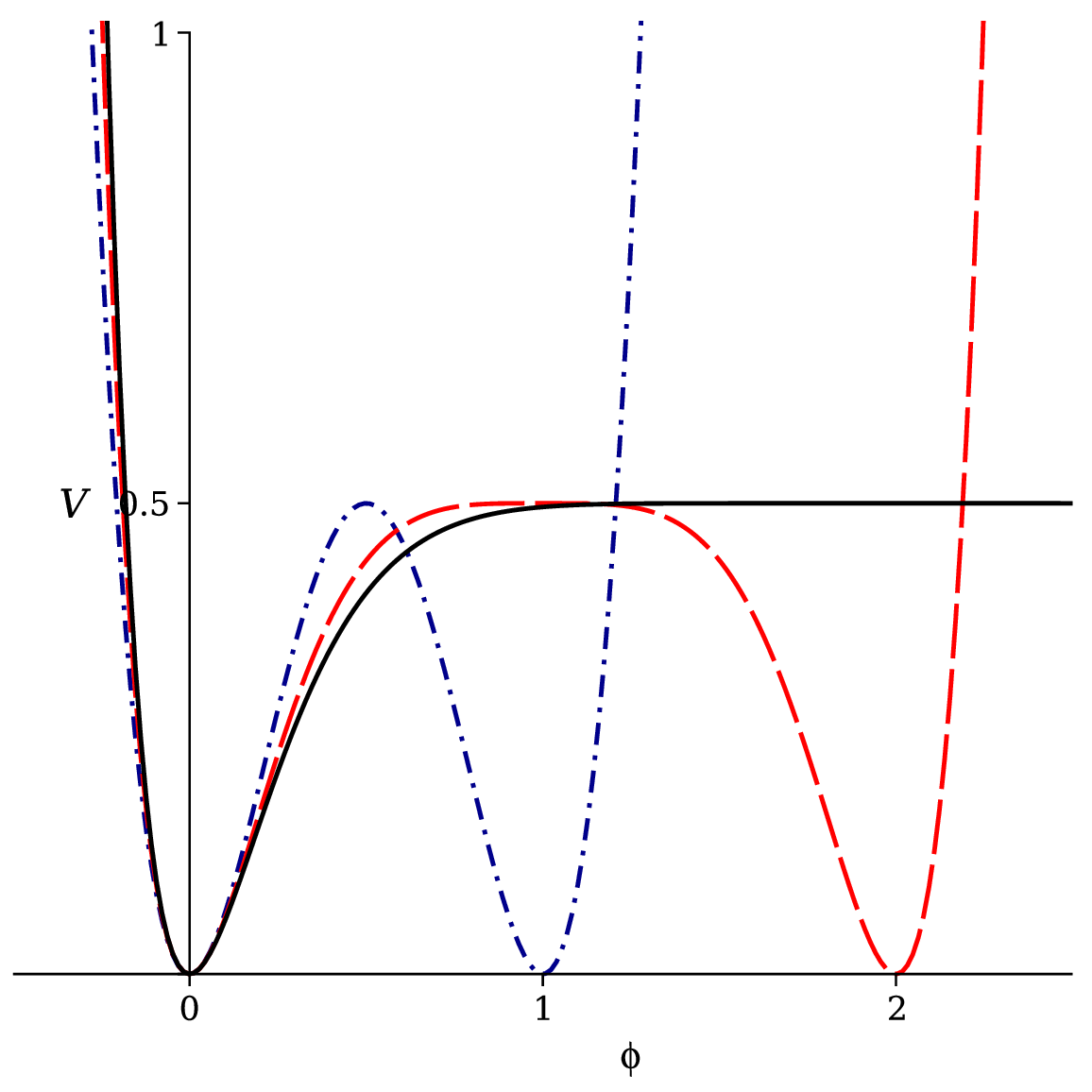}
\includegraphics[scale=0.2]{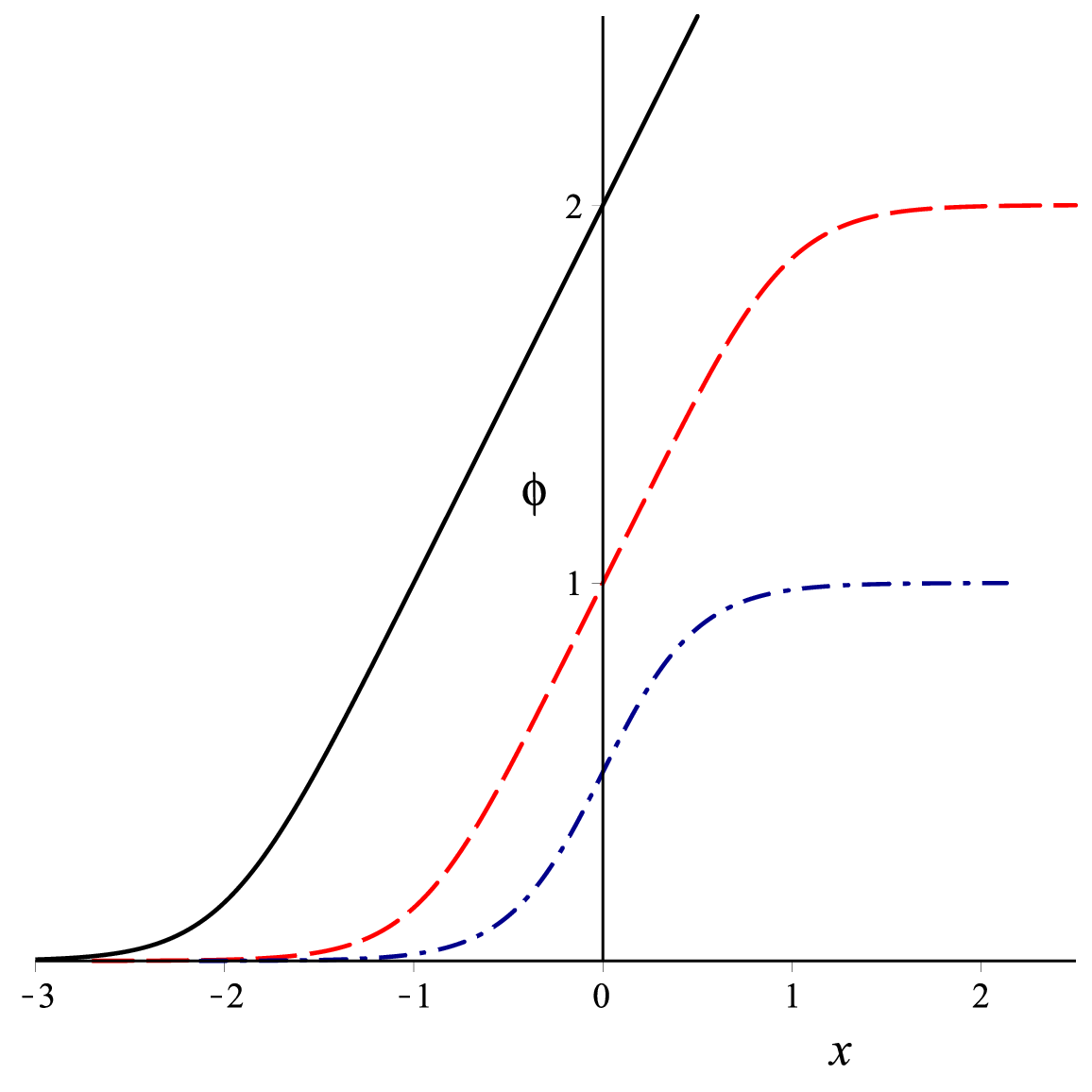}
\includegraphics[scale=0.2]{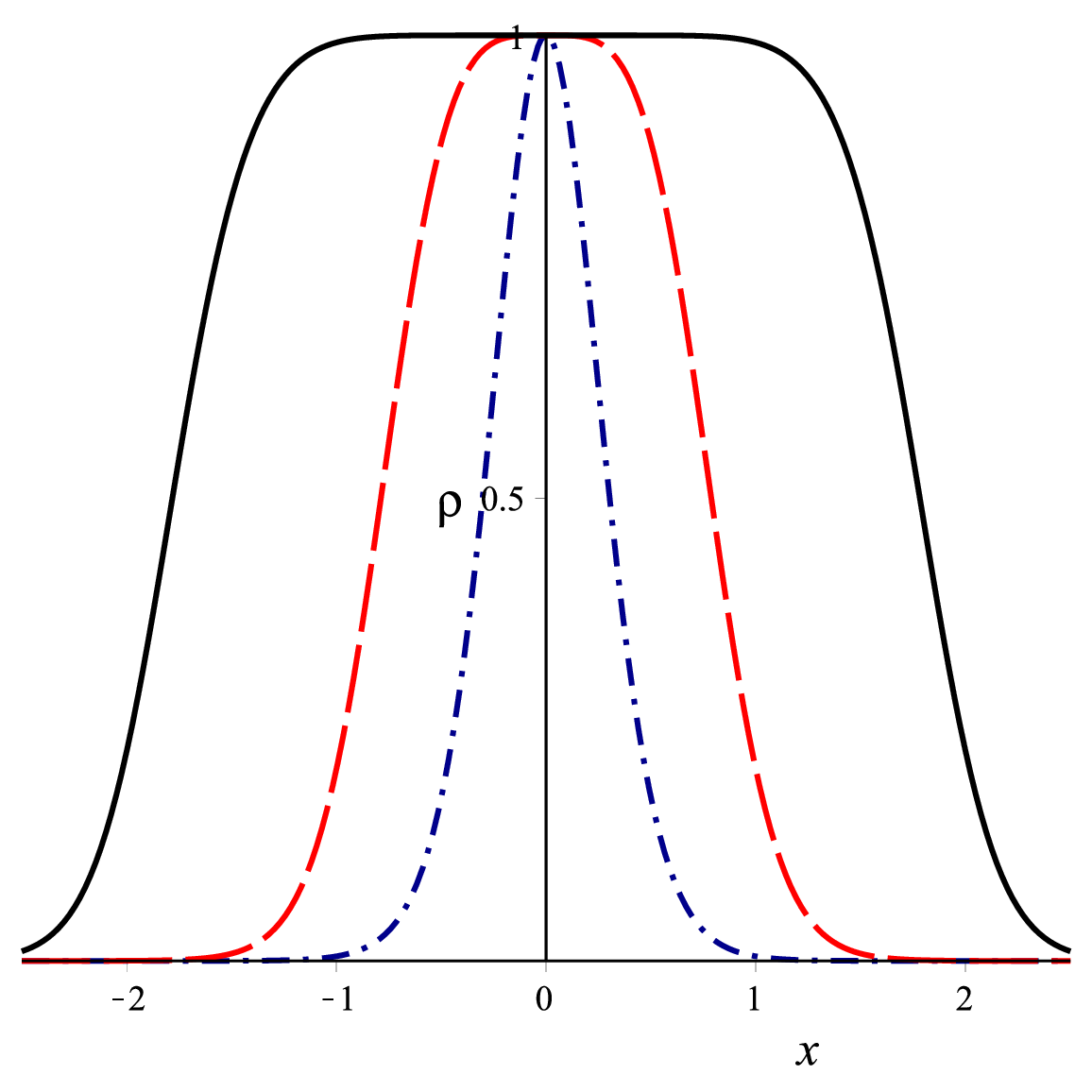}
\includegraphics[scale=0.2]{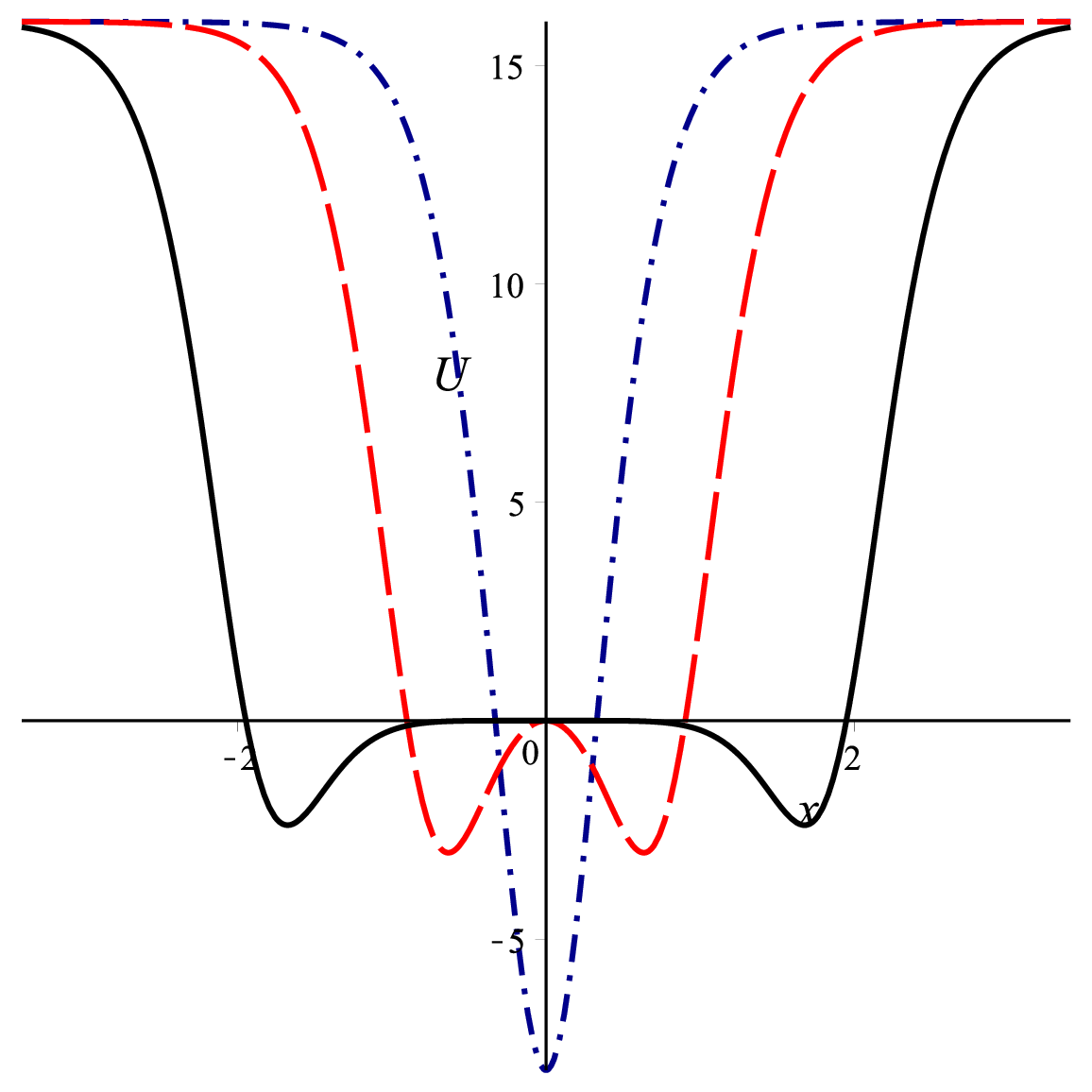}
\caption{The generalized Starobinsky model Eq.~\eqref{Starobinskyg} (first panel), the kink solutions (second panel), the energy densities (third panel) and stability potentials (fourth panel) for $n=1,2,4$, represented by dash-dotted (blue), dashed (red) and solid (black) lines, respectively.}
\label{fig1}
\end{figure}

Now we propose a second deformation
\be
f_2(\phi)=\tanh\left(\frac14 \left(\frac2n\phi-1\right){\bf \Phi}\left(\left(1-\frac2n\phi\right)^{2n},1,\frac1{2n}\right) \right)
\ee
The deformed potential becomes 
\be
V^{(2)}_{n}(\phi)=\frac12\left(1-\left(1-\frac2n\phi\right)^{2n}\right)^2.
\label{Starobinskyg}
\ee 
In terms of the quantities $n=1/2\beta$, $\phi \rightarrow \frac{\phi}{2\sqrt{6}M_{Pl}}$, and $V^{(2)}_{n} \rightarrow V/2V_0$, where $\beta$ is a real parameter, $M_{Pl}$ is the Planck Mass, and $V_0$ is the amplitude of $V(\phi)$, we can write the above potential \eqref{Starobinskyg} as the $\beta-$Starobinsky potential proposed in Ref.~\cite{epjp-2021}, which extends the original Starobinsky model in the following way
\be
V(\phi)=V_0\left(1-\left(1-\sqrt{\frac23}\beta\frac{\phi}{M_{Pl}}\right)^{\frac1\beta}\right)^2.
\label{Vbeta}
\ee
The general framework obtained in \cite{epjp-2021} takes a route based on the brane inflation to investigate observational cosmological
data. Here, we have assumed a different route based on the deformation \cite{Bazeia:2002xg, almeida,bazeialosano} of the theory \eqref{pnormal}, since it was shown in \cite{lima2022} that the generalized Starobinsky can support topological solutions. The  $\beta-$Starobinsky like potential addressed in \eqref{Starobinskyg} presents a topological sector, with two minima and a maximum for each value of $n$, the minima are located at $\phi_{min}=0,n$ and the maximum at $\phi_{max}=n/2$, as can be seen in Fig.~\ref{fig1}.

Although equations \eqref{Vn} and \eqref{Starobinskyg} are somewhat similar, it is important to note that the potential $V^{(2)}_{n}(\phi)$ exhibits a distinct dependence on the parameter $n$ as compared to  $V^{(1)}_{n}(\phi)$. This leads to different potentials, especially when one increases $n$ to higher and higher values. In this situation, the mass does not vary with $n$, and it is given by $m^2=16$. The potential is of the generalized Starobinsky type, supporting kinklike solutions.

Paying further attention to the potential \eqref{Starobinskyg}, we notice that it supports kinklike solutions that obey the expression
\be
\label{solstarstandard}
\left(\frac2n\phi-1\right){\bf \Phi}\left(\left(1-\frac2n\phi\right)^{2n},1,\frac1{2n}\right)=4x.
\ee
They are depicted in the second panel of Fig.~\ref{fig1} for some values of $n$. The energy densities and stability potentials are shown in the third and fourth panels of Fig.~\ref{fig1}, respectively.
 Moreover, using the first-order formalism we can write
\be
W^{(2)}_n=\phi+\frac{n}{2(2n+1)}\left(1-\frac2n\phi\right)^{2n+1}.
\ee
The energy associated to the topological solutions of the $\beta-$Starobinsky like potential $V^{(2)}_{n}(\phi)$ is
\be
E_n^{(2)}=W(n)-W(0)=\frac{2n^2}{2n+1}.
\label{E2}
\ee
As we can see, the relation between equations \eqref{E1}  and \eqref{E2} is given by $E_n^{(2)}=nE_n^{(1)}/2$, and so the energy  $E_n^{(2)}$ diverges as $n$ goes to infinity.

If one considers the expansion for $n\gg1$
\be
\left(1-\frac2n\phi\right)^{2n}=\e^{-4\phi}-\frac4n\phi^2\e^{-4\phi}+{\cal O}\left(\frac{1}{n^2}\right),
\ee
the potential $V^{(2)}_{n}(\phi)$ tends to behave like the original Starobinsky at the limit $n\rightarrow \infty$,
\be
V(\phi)=\frac12\left(1-\e^{-4\phi}\right)^2.
\label{Starobinsky}
\ee 
This asymptotic limit gives rise to the so-called semi-vacuumless potential as described in Refs.~\cite{morris,almeida}, in analogy to the vacuumless potential shown in \cite{cho,dbazeia1999}. For $n$ very large, the potential \eqref{Starobinsky} exhibits one minimum at $\bar\phi_0=0$, which is separated by a barrier from another minimum, $\bar\phi_1$, which moves away from $\bar\phi_0$ and goes to infinity as $n$ increases to larger and larger values. In this situation, the solution diverges asymptotically on the right side. Such behaviour is shown in the upper panels of Fig.~\ref{fig1}.

\begin{figure}%
\centering
\includegraphics[scale=0.3]{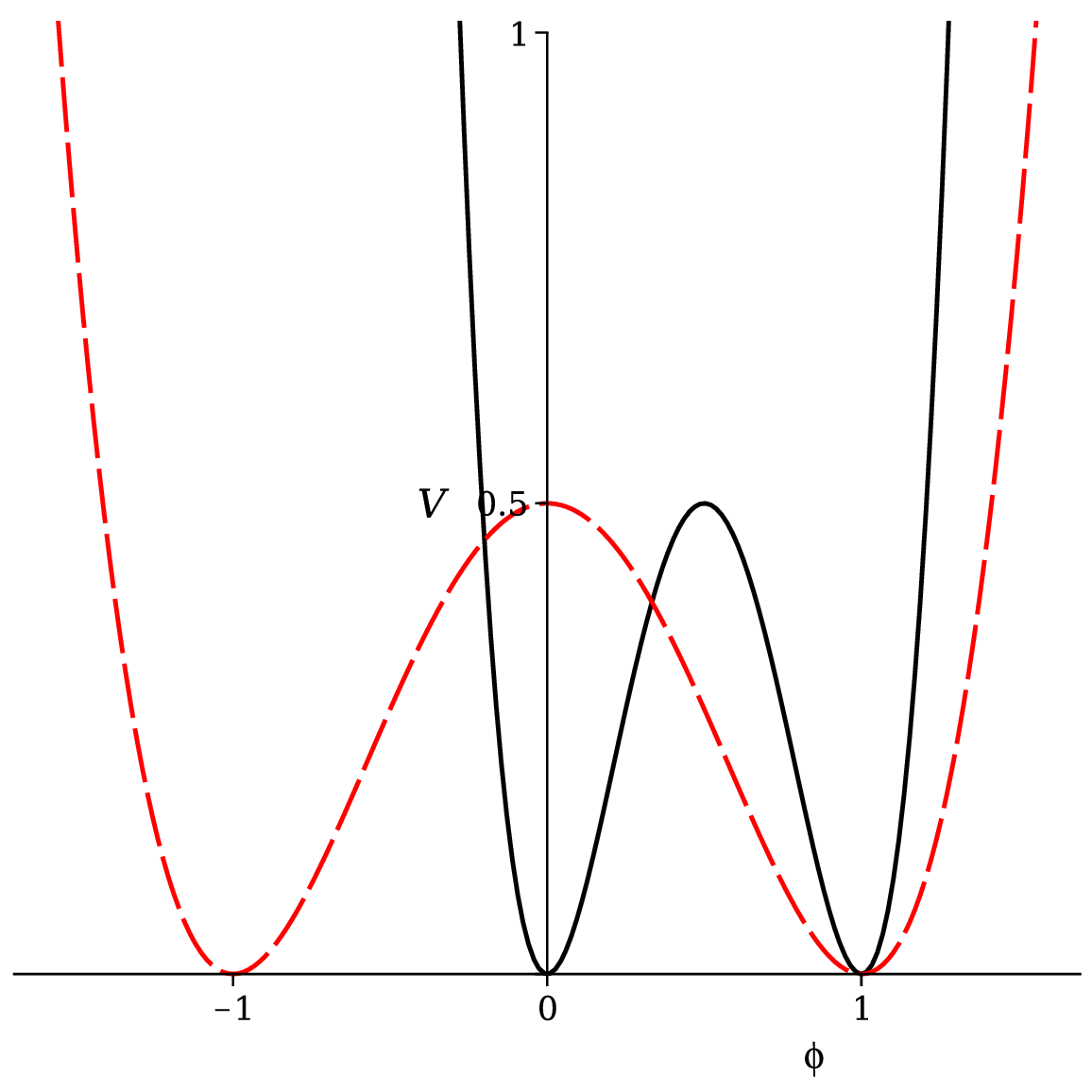}
\caption{The potentials $V_1^{(2)}(\phi)$ given in Eq.~\eqref{v1sta} and $V(\phi)=\frac12(1-\phi^2)^2$, represented by solid (black) and dashed (red) lines, respectively.}
\label{figaa}
\end{figure}

For the particular case $n=1$, the potential \eqref{Starobinskyg} reads
\be
V_1^{(2)}(\phi)=8\phi^2\left(1-\phi\right)^2.
\label{v1sta}
\ee
This model is very similar to the standard $\phi^4$, although the potential \eqref{v1sta}
embodies a shift and shrunk on the $\phi^4$ theory. Fig.~\ref{figaa} depicts the two cases, for comparison. The first-order equation is $\phi'=4\phi(1-\phi)$, and the solution becomes
\be
\label{sol1staro}
\phi(x)=\frac{1 + \tanh(2x)}{2}.
\ee
The energy density is $\rho(x)=\sech^4(2x)$, furnishing the energy $E_1=2/3$. The stability potential is $U(x)=16-24\,\sech^2(2x)$,
which has two bound states with eigenvalues $\omega_0=0$ and $\omega_1^2=12$.

\section{Generalized Kinematics} \label{sec-3}

Generalized models under specific conditions can provide solutions which are compactons \cite{rosenau,compactModels,morecompact}. In this section, we are going to analyze how the defect solutions shown previously respond to a change in their dynamics, and their modification into compact configurations.
\subsection{First generalization} 

In this subsection, we will assume the following extension of the standard dynamics $F(X)=-X^2$. In sequence, we will use the potentials addressed in equations \eqref{Vn} and \eqref{Starobinskyg} to investigate the behaviour of the solutions in this modified scenario. Given the first-order formalism, the potential $V(\phi)$ written in terms of the function $W$ is modified by the dynamics according to $V(\phi)=\frac34W_{\phi}^{4/3}$. Thus, we can rewrite $V^{(1)}_{n}(\phi)$ as
\be
V^{(1)}_n(\phi)=\frac{3}{4}\left(1-\phi^{2n}\right)^2,
\label{Vnm}
\ee
and the first-order equation $\phi'=W_{\phi}^{1/3}$ becomes
\be
\phi'=\sqrt{1-\phi^{2n}}.
\label{p1m}
\ee

The solution is given in terms of the ordinary hypergeometric function ${\bf _2F_1}$
\be
\label{solVnX}
\phi\; {\bf _2F_1}\left(\frac12,\frac1{2n};1+\frac1{2n};\phi^{2n}\right)= x,
\ee
and the minima $\bar\phi_{\pm}=\pm1$ are reached exactly at the points $\bar{x}_\pm=\pm \left.\pi^{1/2}\Gamma\left(\frac{2n+1}{2n}\right)\right/\Gamma\left(\frac{n+1}{2n}\right)$. In this case, for each value of $n$, the defect solution \eqref{solVnX} lives in a compact region between $\bar{x}_{-}$ and $\bar{x}_{+}$. The energy density is null outside the compact region. This differs from the standard situation \eqref{solVnstandard} where the compact regime was only gotten for $n$ very large.

\begin{figure}%
\centering
\includegraphics[scale=0.2]{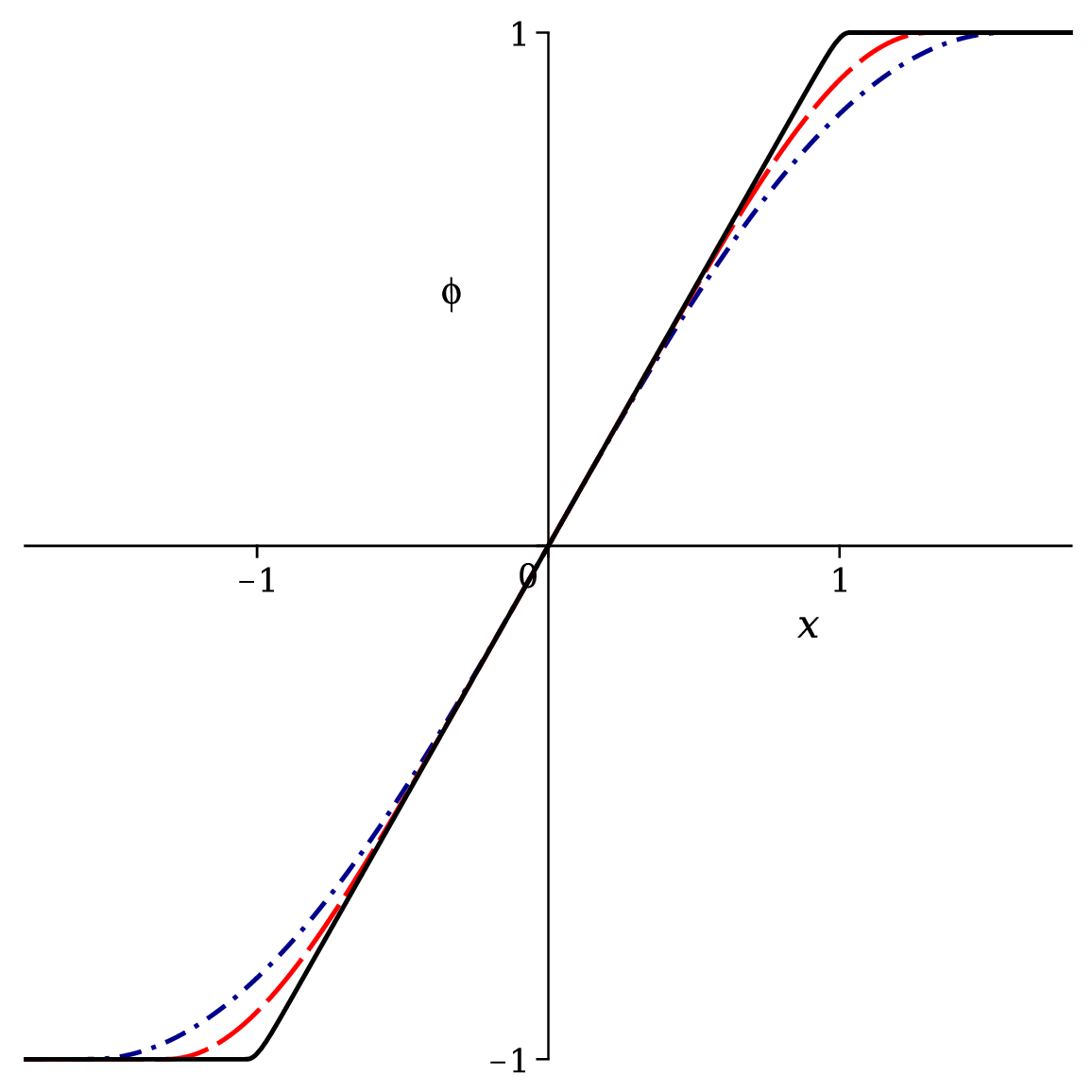}
\includegraphics[scale=0.2]{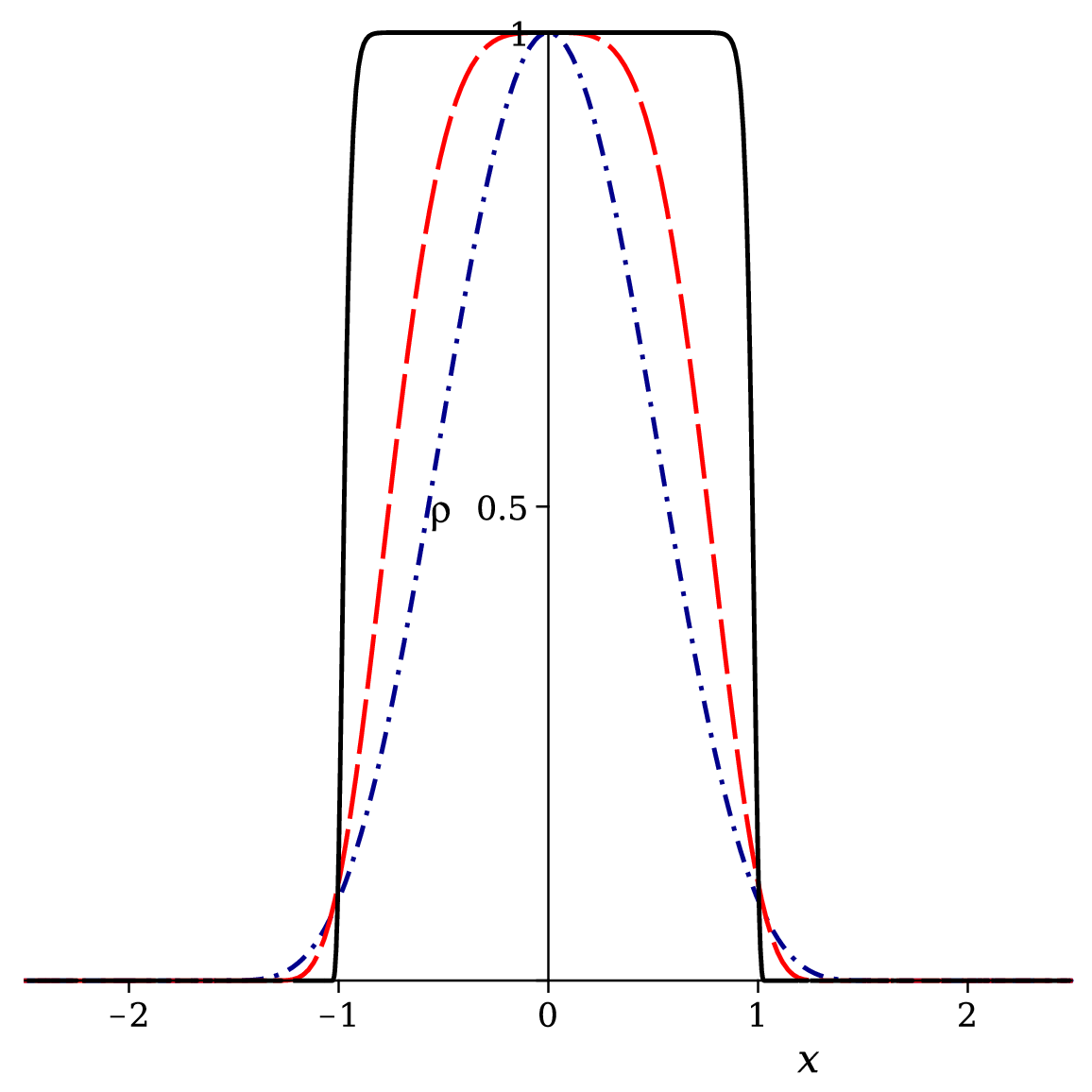}
\caption{The solutions \eqref{solVnX} (first panel), and the energy densities (second panel) in the first modified scenario, for the same values of $n$ adopted in Fig.~\ref{fig1A}.}
\label{fig2}
\end{figure}

Fig.~\ref{fig2} displays the solution \eqref{solVnX} and the energy density for increasing values of $n$. The space interval where the solution is trapped gets smaller as $n$ increases until it reaches the values $\bar{x}_\pm=\pm1$. The largest value of $\bar{x}_+$ is gotten when $n=1$ ($\bar{x}_+=\pi/2$) and the smallest one is reached when $n\rightarrow \infty$  ($\bar{x}_+=1$).

It is important to observe how the standard compacton behaves in this generalized scenario, by taking the limit $n\rightarrow \infty$. The hypergeometric function given in \eqref{solVnX} becomes the identity, ${\bf _2F_1}\left(\frac12,0;1;\phi^{2n}\right)=1$. In this case, the solution and energy density are equivalent to the ones found in the standard case.  Interestingly, this generalized model does not change the standard compact profile, which appears to be a twin of the standard model, supporting the same defect structure with the same energy density, as studied before in Refs. \cite{T1,T11,T2,T111,T3,T4,T5,T6}.

 For the case $n=1$, the solution becomes
\begin{eqnarray}\label{sinx}
\phi(x) = \left\{
\begin{array}{ll}
-1, \qquad \mbox{for} \quad \, x < -\pi/2,\\ 
 \sin(x), \,\, \,\,\mbox{for} \quad |x| \leq \pi/2,\\
1, \qquad \,\,\,\, \,\mbox{for}\quad \, x > \pi/2\,,
\end{array} \right.
\end{eqnarray}
The energy density inside the compact interval is $\rho(x)=\cos^4(x)$, and null elsewhere. The stability potential is of the P\"oschl-Teller type  $U(z)=-12+6\sec^2\left(\sqrt{3}z\right)$ for $|z| \leq \pi/2\sqrt{3}$, and infinite outside this region. The states are bounded with eigenvalues of energy given by $\omega_{k}^2=12k(k+2)$, where $k=0,1,2,\cdots,\infty$.

The second potential investigated is written as
\be
V_{n}^{(2)}(\phi)=\frac34\left(1-\left(1-\frac2n\phi\right)^{2n}\right)^2,
\label{Starobinskym}
\ee 
that furnishes the first-order equation
\be
\phi'=\sqrt{1-\left(1-\frac2n\phi\right)^{2n}}.
\ee
Here, the solutions also assume a compact behaviour for each value of $n$. The potential minima $(\phi_{min}=0,n)$  are reached  at finite points of $x$, which are $$\bar{x}_\pm=\pm\left.n\pi^{1/2}\Gamma\left(\frac{2n+1}{2n}\right)\right/2\Gamma\left(\frac{n+1}{2n}\right).$$

\begin{figure}%
\centering
\includegraphics[scale=0.2]{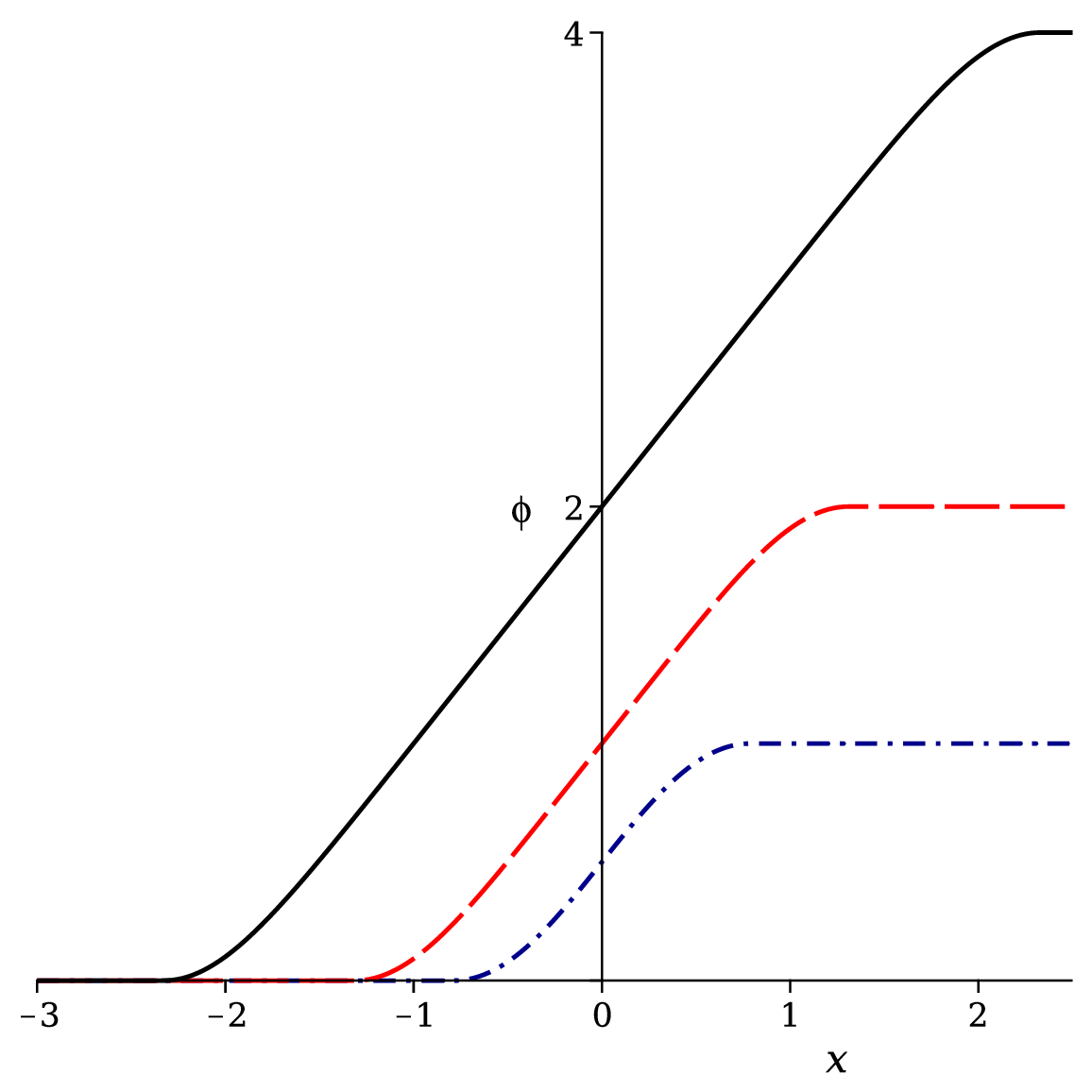}
\includegraphics[scale=0.2]{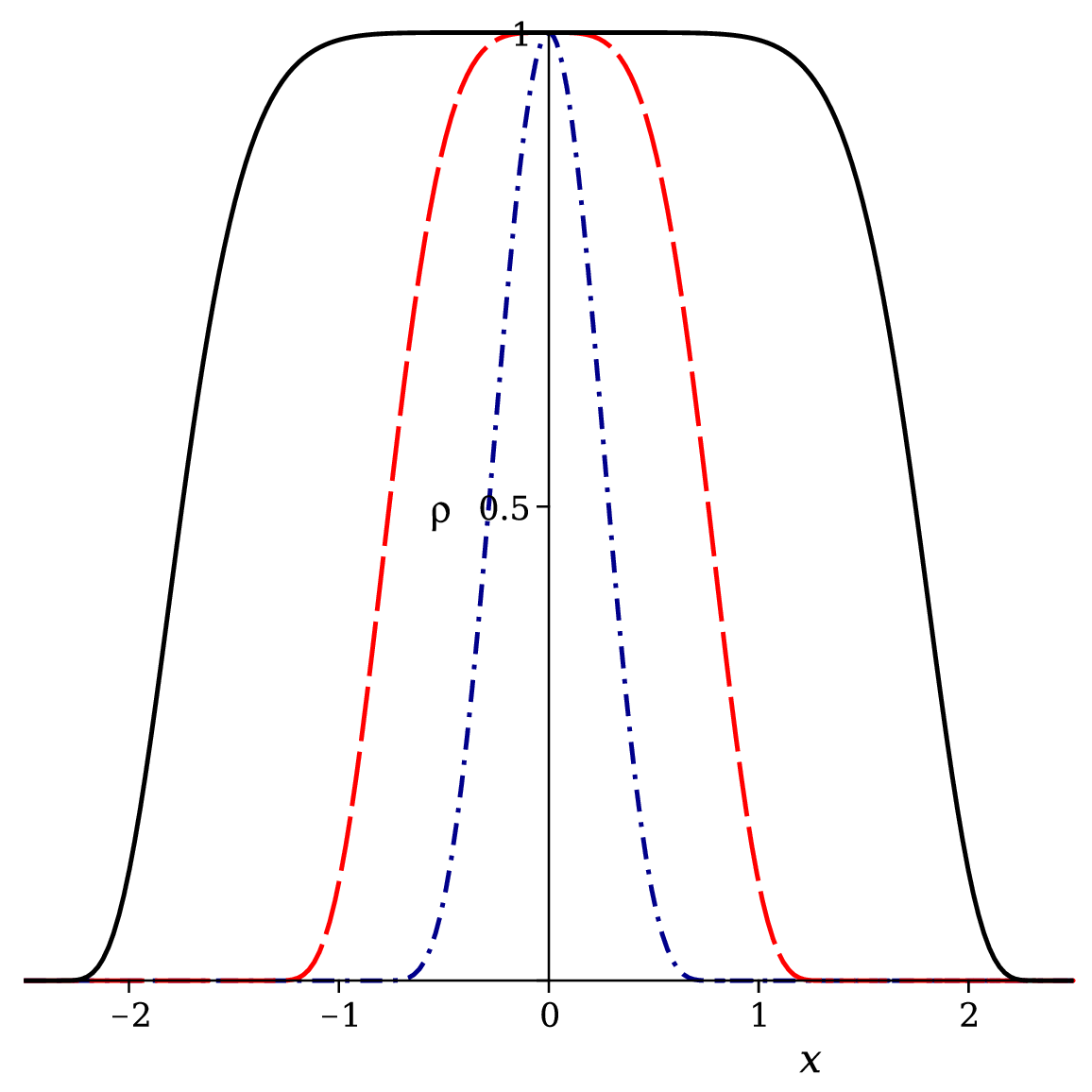}
\caption{The first panel shows the solution \eqref{solstaromod} and the second panel the energy density in the first modified scenario, using the same values of $n$ adopted in Fig.~\ref{fig1}. }
\label{fig3}
\end{figure}

The compacton solution is given by the relation
\be
\label{solstaromod}
\left(1-\frac2n\phi\right)\;  {\bf _2F_1}\left(\frac12,\frac1{2n};1+\frac1{2n};\left(1-\frac2n\phi\right)^{2n}\right)= - \frac2n x,
%
%
\ee
for $x\in \left[\bar{x}_-,\bar{x}_+\right]$, $\phi(x)=0$ for $x< \bar{x}_{-}$, and $\phi(x)=n$ for $x> \bar{x}_+$. The solution above and its energy density are shown in Fig.~\ref{fig3} for some values of $n$.

The solution for $n=1$ is very similar to the one obtained in \eqref{sinx}
\begin{eqnarray}
\label{solstarcompacton}
\phi(x) = \left\{
\begin{array}{ll}
0, \qquad   \quad \quad \,\mbox{for} \quad  \, x < -\pi/4,\\ 
\frac{1+\sin(2x)}{2}, \quad \mbox{for} \quad   |x| \leq \pi/4,\\
1,  \qquad \quad \quad \, \mbox{for} \quad  \,x > \pi/4\,,
\end{array} \right.
\end{eqnarray}
The energy density becomes $\rho(x)=\cos^4(2x)$ for $ |x| \leq \pi/4$, and null elsewhere. The stability potential is $U(z)=-48+24\sec^2\left(2\sqrt{3}z\right)$ for $|z| \leq \pi/4\sqrt{3}$, and infinity outside this interval. It has infinite bound states, for  $k=0,1,2,\cdots$, the corresponding eigenvalues of energy are $\omega_{k}^2=48k(k+2)$. 

\begin{figure}%
\centering
\includegraphics[scale=0.2]{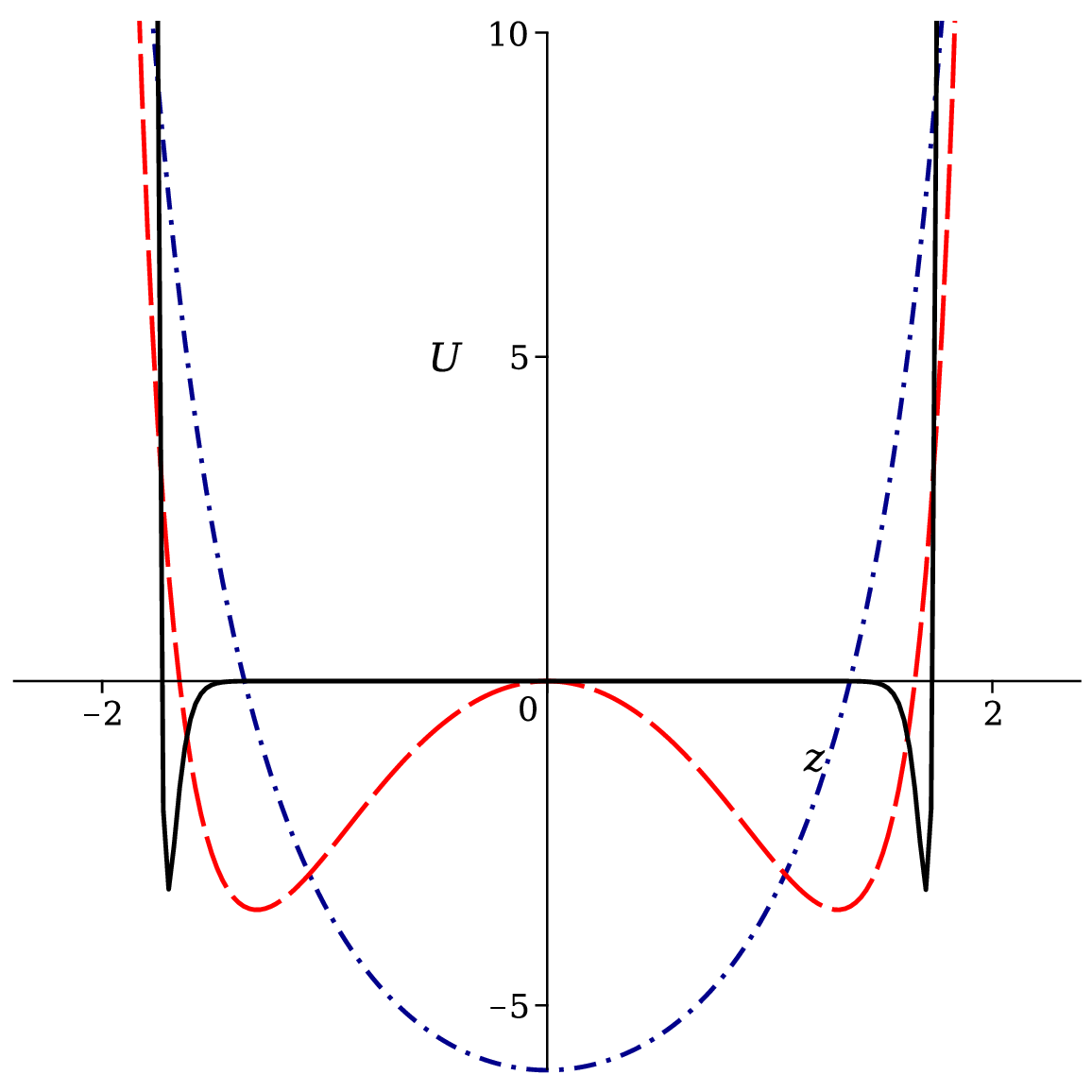}
\includegraphics[scale=0.2]{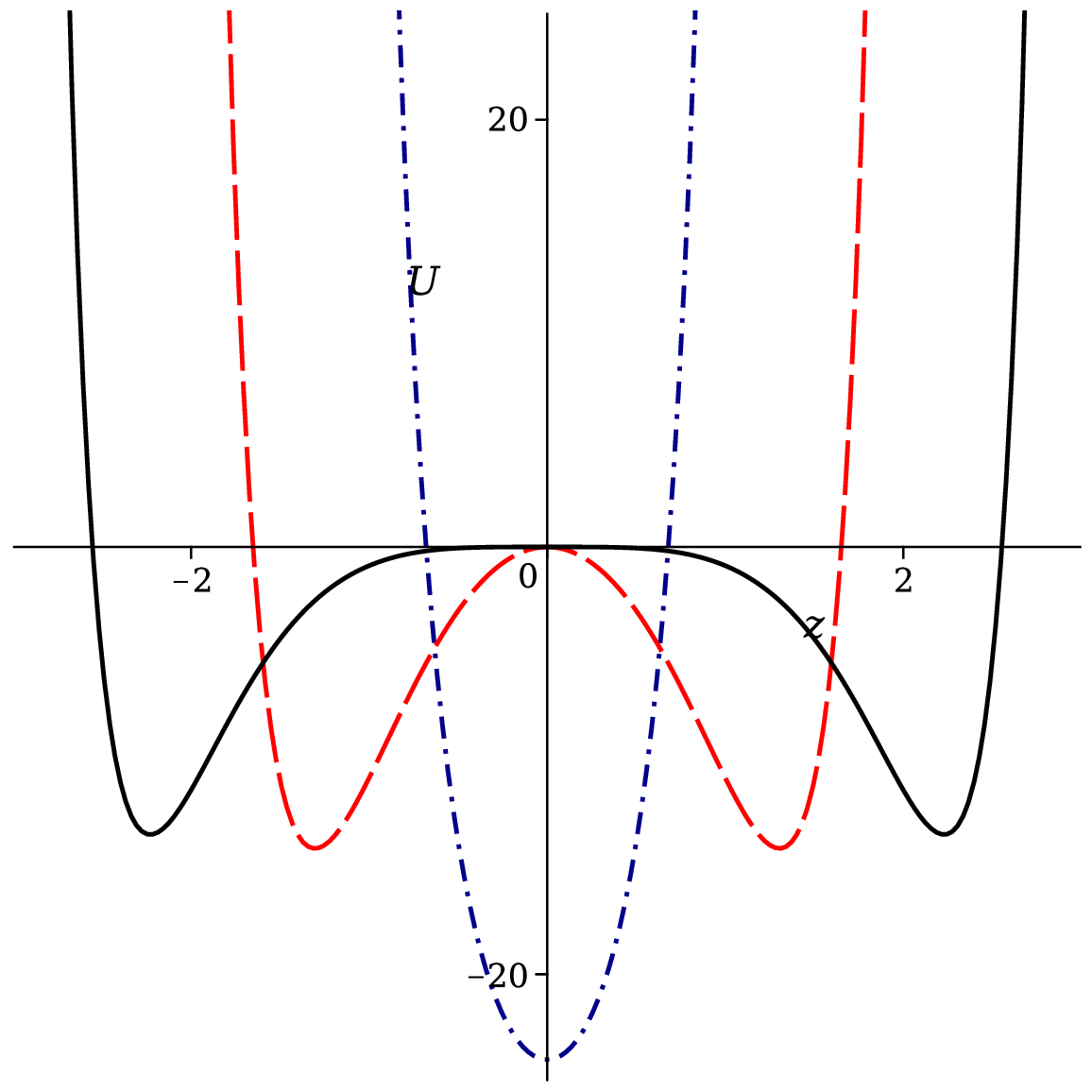}
\caption{The stability potentials in the generalized case. The first panel shows  $U(z)/n^2$ for the first model \eqref{Vnm} with $n=1,2,20$. The second panel shows $U(z)$ for the $n-$Starobinsky model \eqref{Starobinskym} with  $n=1,2,4$.}
\label{fig2A}
\end{figure}

Fig.~\ref{fig2A} represents the stability potentials in this generalized case, comparing the results gotten from the $n-$Starobinsky \eqref{Starobinskym} and the first model \eqref{Vnm}. Both situations are very alike for lower values of $n$, but they become very different as $n$ increases.

\subsection{Second generalization} \label{sec4b}

As we have seen, the standard kink solutions analyzed evolve to compactons when the dynamic is altered for a quadratic form. To explore how this transition occurs, one uses a second extension of the standard kinetics, given by $F(X)=X-\alpha X^2$, where the parameter $\alpha$ controls the modification that leads from a standard to a generalized scenario. Although it is expected that in the limit $\alpha \rightarrow 0$, the models return to the standard case with $F=X$, and for $\alpha \gg 1$, they lead to the generalized case with $F=-X^2$, the introduction of $\alpha$ modifies the standard potentials $V^{(1)}_n$ and $V^{(2)}_n$, embedding them into a wider context.  This allows for a deeper understanding of the solution behaviour and its properties as $\alpha$ increases.

The equation of motion for static fields is
\be
(1+3\alpha\phi'^2)\phi''=V_\phi.
\ee
It can be reduced to first-order by integration
\be
\frac{1}{2}\phi'^2+\frac34\alpha\phi'^4=V,
\ee
which also can be written as
\be
\frac{1}{2}\phi'^2=V_{mod}(\phi),
\label{foetransit}
\ee
where
\be
V_{mod}(\phi)=\frac{1}{6\alpha}\left(\sqrt{1+12\alpha V}-1\right).
\ee
The energy density for this extension becomes
\be
\rho=\phi'^2+\alpha\phi'^4.
\ee

\begin{figure}%
\centering
\includegraphics[scale=0.3]{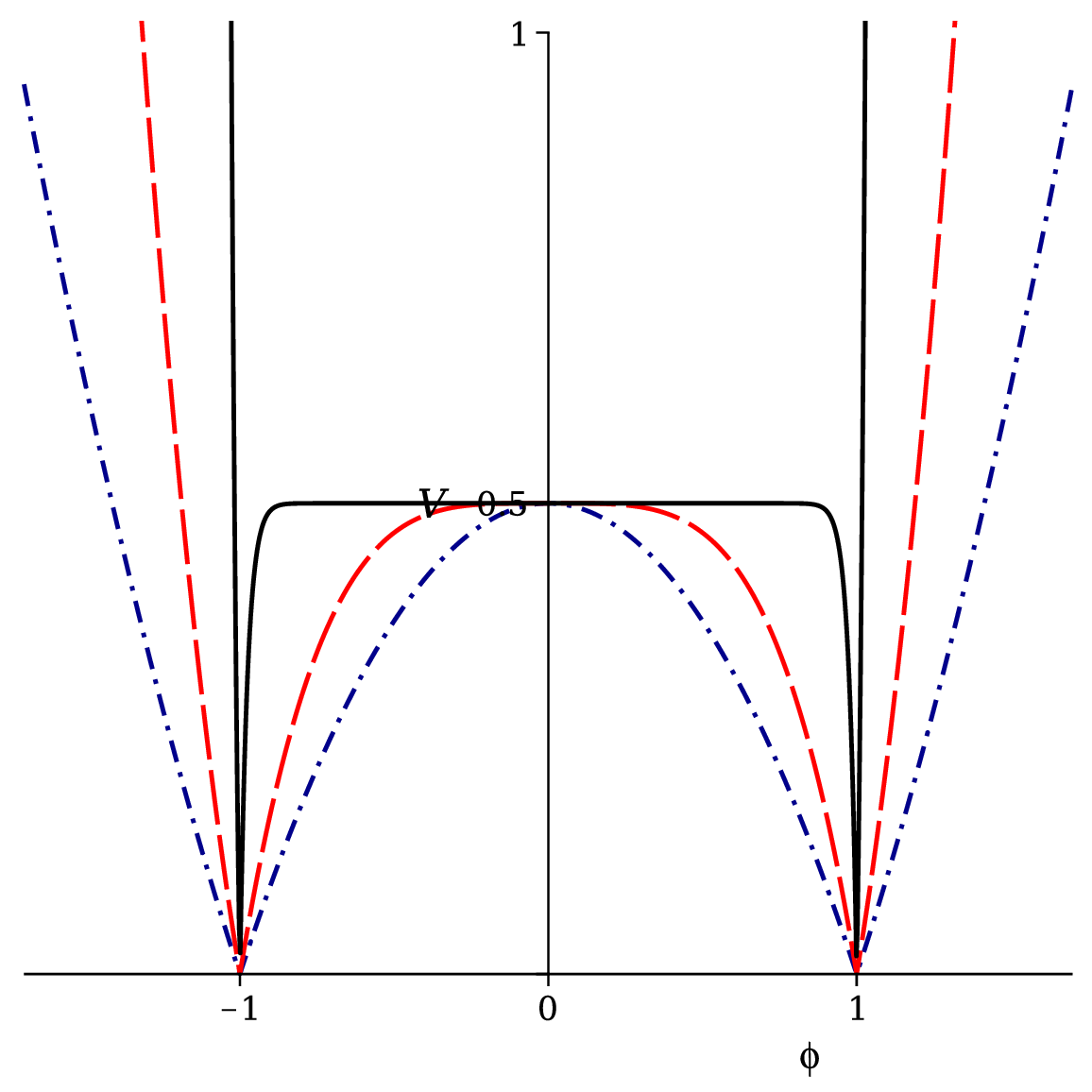}
\caption{The potential \eqref{moduloVn}, depicted for $n=1, 2, 20$ with dash-dotted (blue), dashed (red) and solid (black) lines, respectively.}
\label{fig5}
\end{figure}

The minima given by $V_{mod}(\phi_{min})=0$ implies in $V(\phi_{min})=0$, so the distance between the minima does not depend on $\alpha$.
However, the height of the potential $V_{mod}(\phi)$ at the maximum point varies with $\alpha$, and the width of the defect gets a strong dependence on $\alpha$. To avoid that dependence and help us to understand better the defect structure for large $\alpha$, we modify the potential $V(\phi)$ in $V_{mod}(\phi)$, just as $V_{mod}(\phi_{max})=1/2$. Then,
\be
V(\phi)\rightarrow \left(1+\frac32\alpha\right)V_s(\phi)
\ee
where $V_s(\phi)$ is the potential in the standard scenario; and the generalized potential becomes 
\be
V_{mod}(\phi)=\frac{1}{6\alpha}\left(\sqrt{1+12\alpha \left(1+\frac32\alpha\right)V_s(\phi)}-1\right).
\label{Vmod}
\ee
For $\alpha$ very small, we get
\be
V_{mod}(\phi) =V_s+\frac{3\alpha}{2}V_s\left({1-2V_s}\right) +{\cal O}\left({\alpha^2}\right).
\label{smallalpha}
\ee
For the regime $\alpha$ very large, we make the expansion $1/\alpha \ll 1$, so we get
\be
V_{mod}(\phi) =\sqrt{\frac{V_s}{2}}+\frac{1}{6\alpha}\left({\sqrt{2V_s}-1}\right) +{\cal O}\left(\frac{1}{\alpha^2}\right).
\label{bigalpha}
\ee

The first standard potential selected  is $V_s(\phi)=V^{(1)}_n(\phi)$ given in \eqref{Vn}. That furnishes a most general model which depends on both parameters $\alpha$ and $n$
\be
V^{(1)}_{mod}(\phi)=\frac{1}{6\alpha}\left(\sqrt{1+6\alpha \left(1+\frac32\alpha\right)\left(1- \phi^{2n}\right)^2}-1\right).
\label{v1mod}
\ee
The modified mass is $m^2_{mod}=4n^2\left(1+3\alpha/2\right)$, which increases to large values when $n$ or $\alpha$ get larger.  
To comprehend the behaviour of $V^{(1)}_{mod}(\phi)$ for different values of parameters, let us examine the extreme regimes $\alpha$ very small and   $\alpha$ very large. In the limit $\alpha\rightarrow 0$, the modified potential recovers the standard one \eqref{Vn}. The mass goes with $m^2=4n^2$; the solution is given in \eqref{solVnstandard}, which leads a kink into a compacton as $n$ assumes increasing values. 

\begin{figure}%
\centering
\includegraphics[scale=0.2]{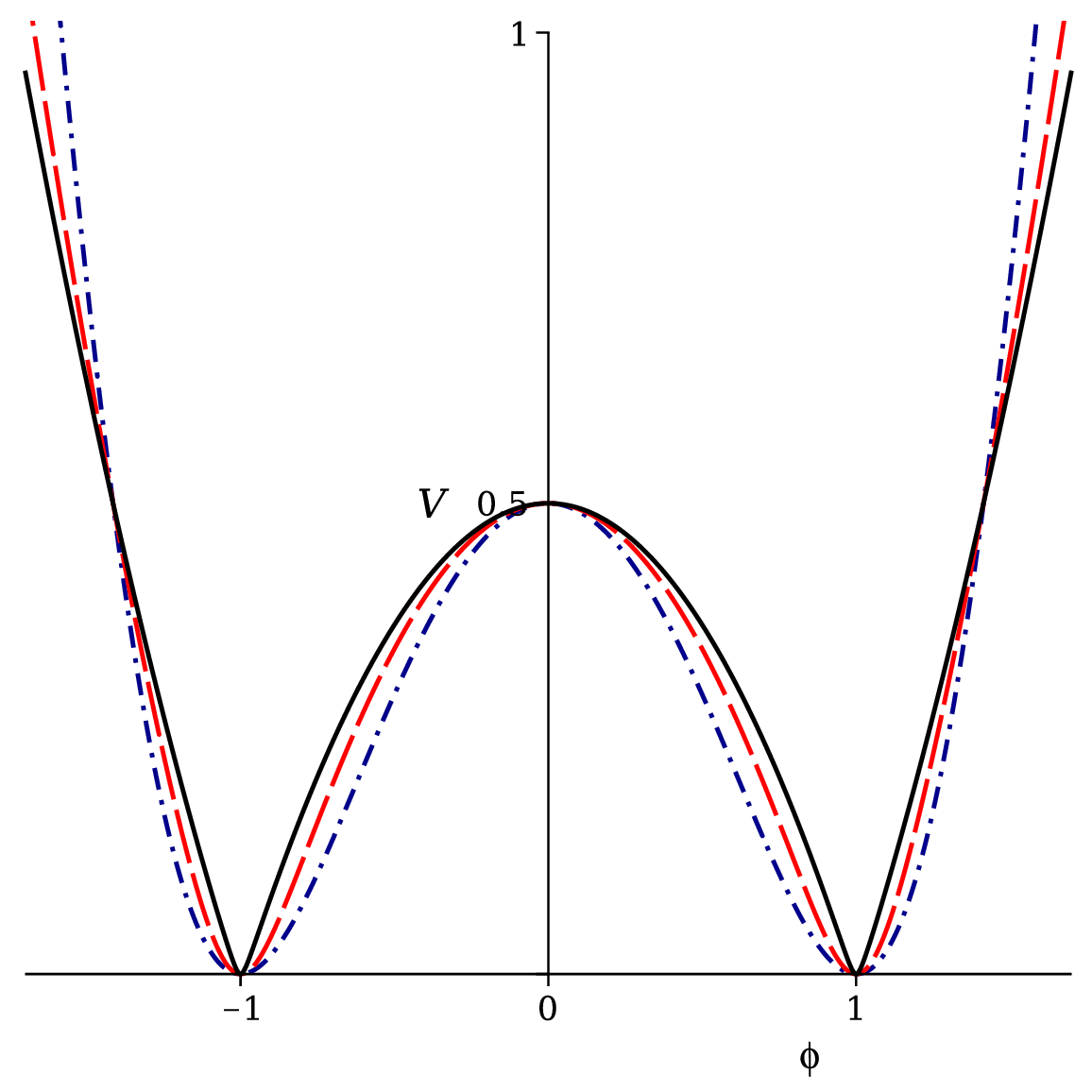}
\includegraphics[scale=0.2]{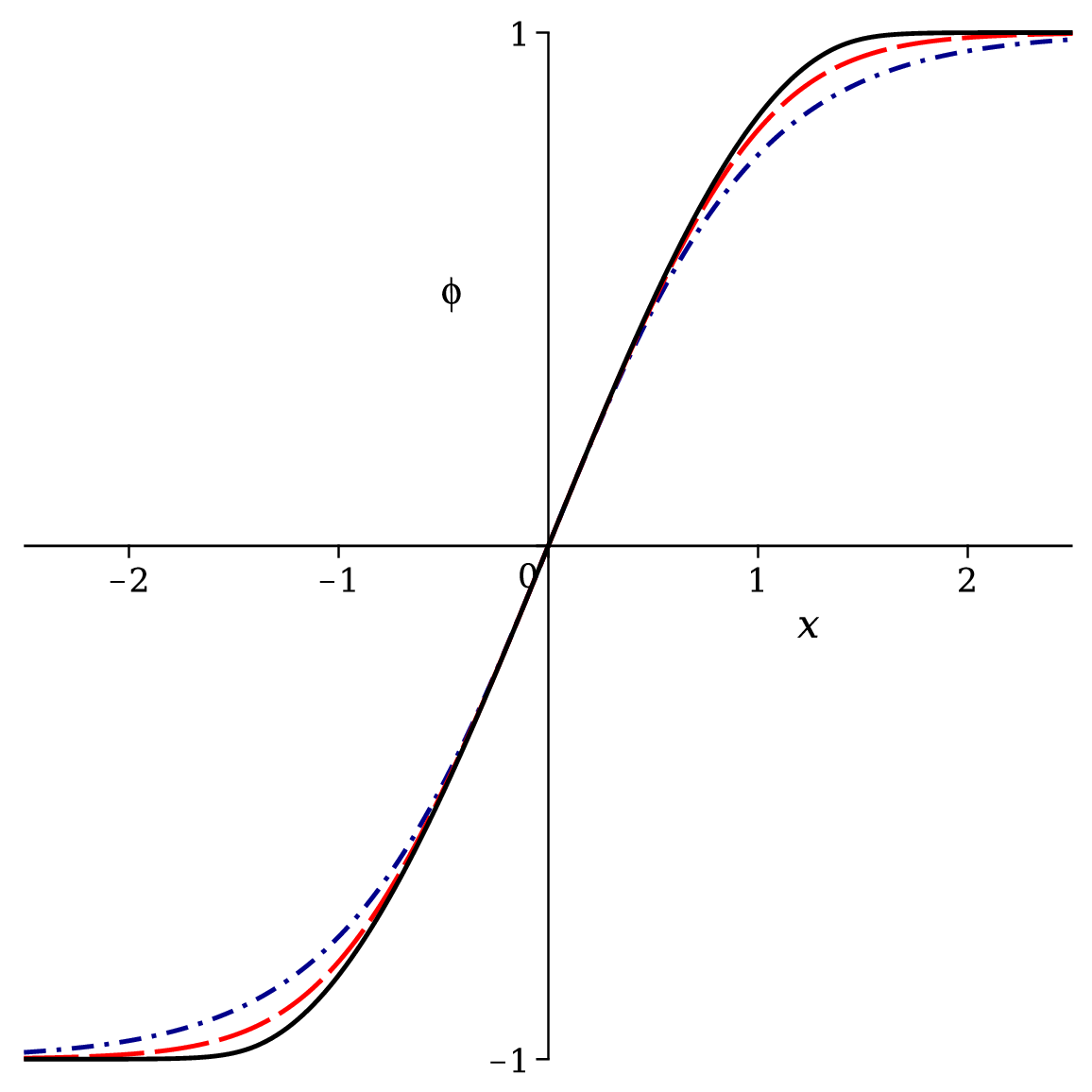}
\caption{The first panel shows the potential \eqref{v1mod} and the second panel shows the solution, for $n=1$ and $\alpha=0, 1, 10$ depicted by dashed-dot (blue), dashed (red) and solid (black) lines, respectively.}
\label{fig4}
\end{figure}

In contrast, taking the limit $\alpha\rightarrow \infty$, the expression for $V^{(1)}_{mod}(\phi)$ becomes
\be
V_{mod}^{(1)}(\phi)=\frac12|1-\phi^{2n}|,
\label{moduloVn}
\ee
which provides a set of compact solutions expressed as the same ones shown in Eq.~\eqref{solVnX}, see first panel on Fig.~\ref{fig2}. The solution is of the compacton type for each value of $n$, and the potential \eqref{moduloVn} is depicted in Fig.~\ref{fig5}.

These results suggest that the solution can transit from the standard case \eqref{solVnstandard} to the generalized \eqref{solVnX} by varying the parameter $\alpha$, for any value of $n$. In the case $n=1$, the generalized potential \eqref{v1mod} becomes identical to another specific scenario studied in Ref.~\cite{menezes2014}. The mass varies according to $m^2_{mod}=4+6\alpha$. At the limit $\alpha=0$, the solution is the kink of the standard $\phi^4$ theory; and at the limit $\alpha\rightarrow \infty$, the solution takes the compact form given by the same expression shown in  Eq.~\eqref{sinx}. In this circumstance, the modified potential \eqref{v1mod} transforms the solution kink into compacton for increasing values of $\alpha$. This behaviour is depicted in Fig.~\ref{fig4} and the energy densities are shown on the first panel of Fig.~\ref{fig8a} for some values of $\alpha$.
However, considering the regime $n \gg 1$ in \eqref{v1mod}, the solution profile does not change with $\alpha$. The modified potential is such that $V_{\alpha\rightarrow 0}=(1-\phi^{2n})^2/2$ and $V_{\alpha\rightarrow\infty}=|1-\phi^{2n}|/2$. They tend to behave in the same way when $n$ is very large. In both situations, the solution is the compacton $\phi(x)=x$ for $x\in[-1,1]$, $\phi(x)=-1$ for $x<-1$, and $\phi(x)=1$ for $x>1$. The transition from the standard dynamics to the generalized one occurs in a way that, when $n$ is very large, the solution and energy density do not change.

From now on, we are going to use the standard Starobinsky like potential given in \eqref{Starobinskyg}, $V_s(\phi)=V^{(2)}_n(\phi)$. The expression for $V_{mod}(\phi)$ reads
\be
\footnotesize 
V_{mod}^{(2)}(\phi)=\frac{1}{6\alpha}\left(\sqrt{1+3\alpha \left(2+3\alpha\right)\left(1-\left(1-\frac{2\phi}n\right)^{2n}\right)^2}-1\right).
\label{Vmod2}
\ee
The second modified potential $V_{mod}^{(2)}(\phi)$ inserts the generalized Starobinsky  into a broader framework, where $V^{(2)}_n(\phi)$ is recovered at the limit $\alpha\rightarrow 0$.  The modified mass is $m^2_{mod}=16+24\alpha$. 

\begin{figure}%
\centering
\includegraphics[scale=0.3]{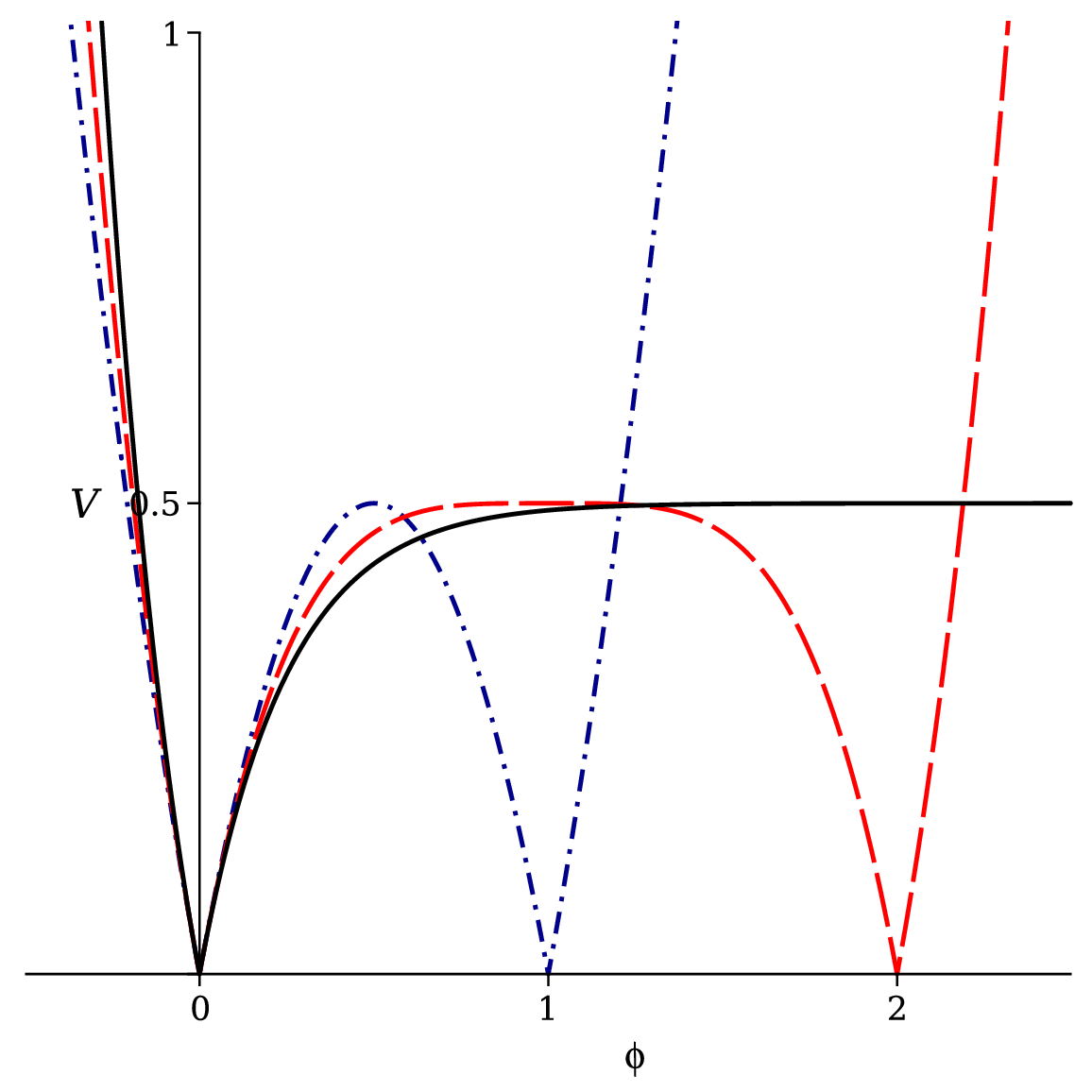}
\caption{The potential \eqref{moduloStar} for $n=1, 2, 4$ depicted by dashed-dot (blue), dashed (red) and solid (black) lines, respectively.}
\label{fig7}
\end{figure}

The limit $\alpha\rightarrow \infty$ gives
\be
\small
V_{mod}^{(2)}(\phi)=\frac12\left|1-\left(1-\frac{2\phi}n\right)^{2n}\right|.
\label{moduloStar}
\ee
The solution at this regime becomes of the compact type for each value of $n$, being given by Eq.~\eqref{solstaromod},  see the first panel in Fig.~\ref{fig3}. The behaviour of the potential \eqref{moduloStar} is represented in Fig.~\ref{fig7}.

\begin{figure}%
\centering
\includegraphics[scale=0.2]{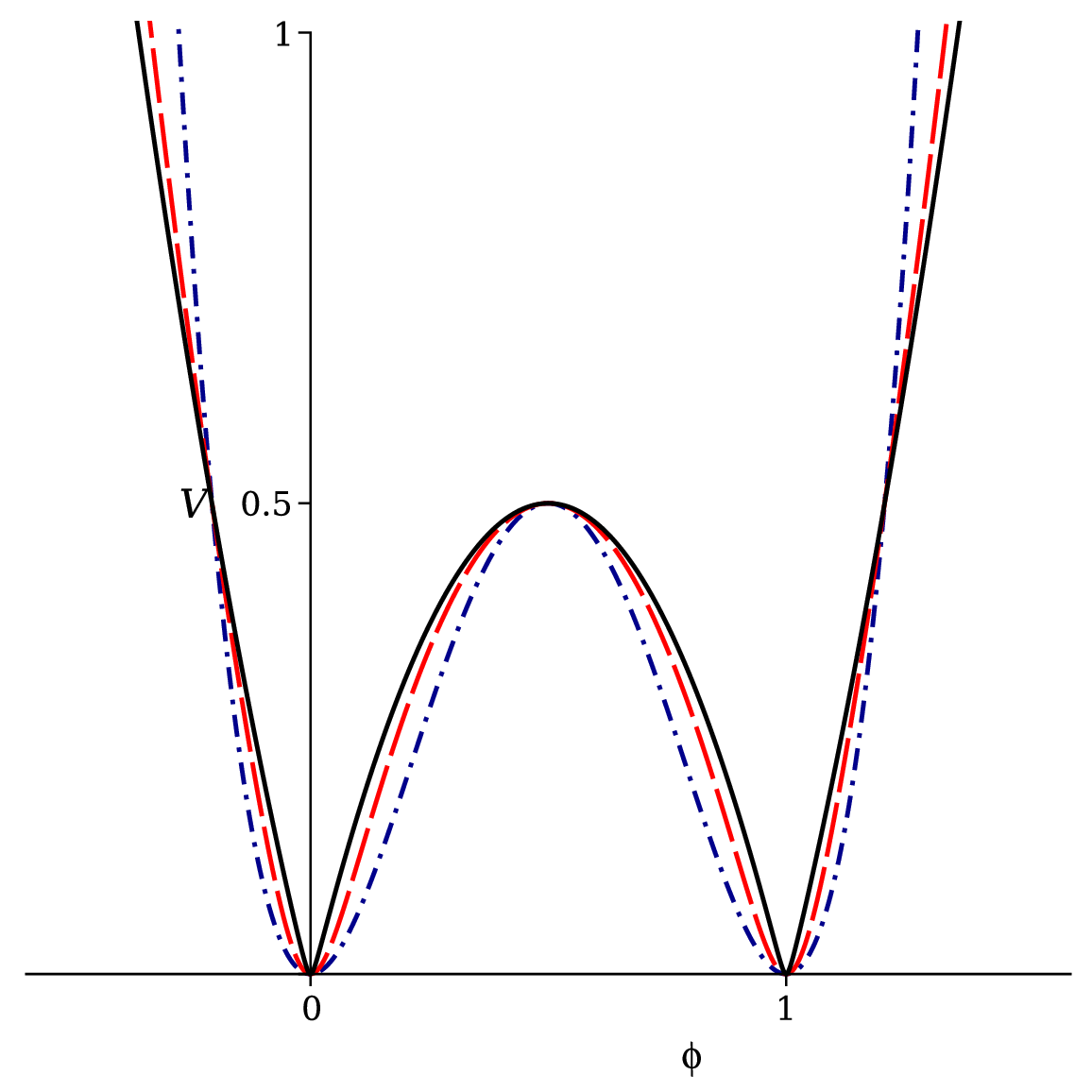}
\includegraphics[scale=0.2]{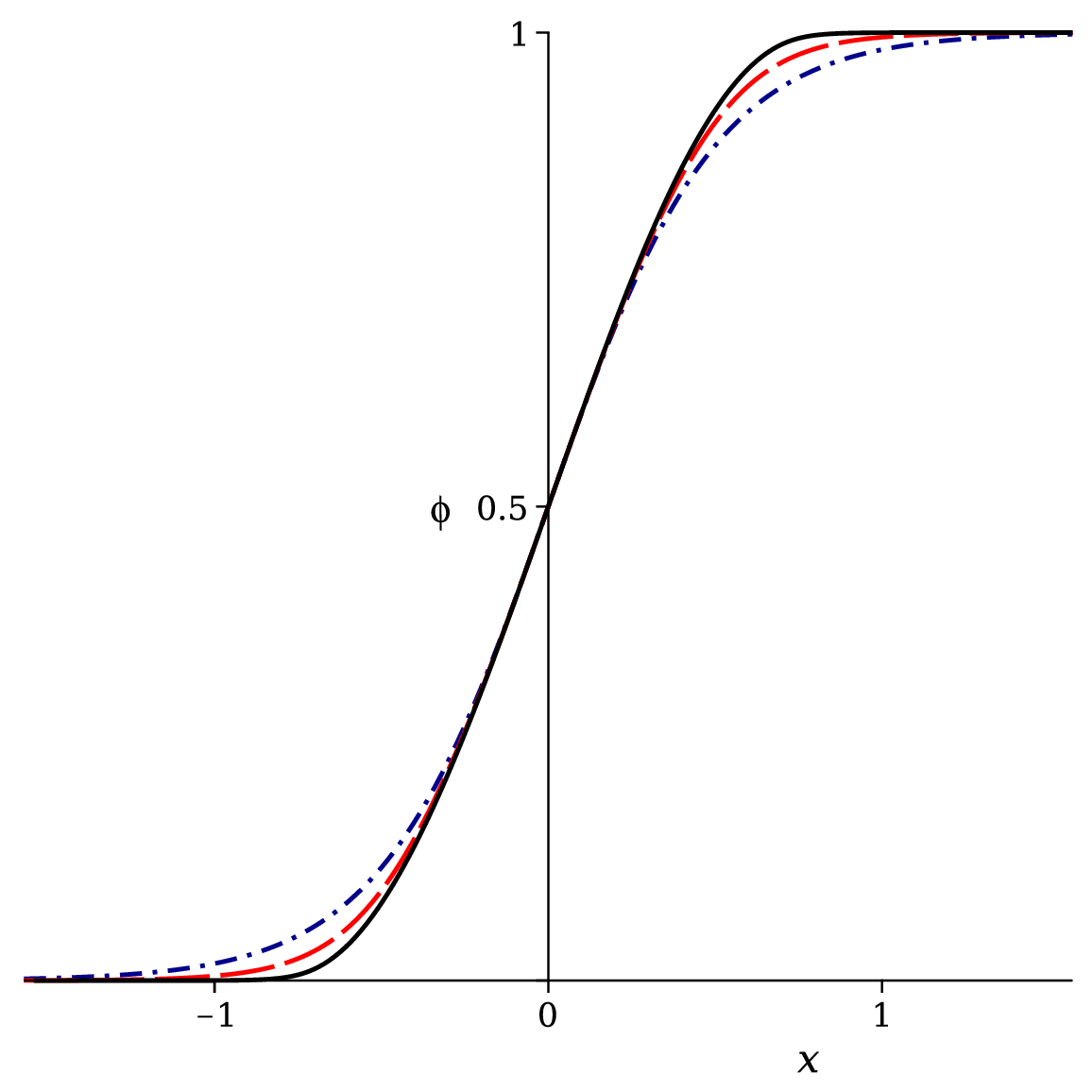}
\caption{First panel shows the potential \eqref{Vmod2} and second panel shows the solution going from \eqref{sol1staro} to \eqref{solstarcompacton}, for $n=1$ and $\alpha=0,1,10$ depicted by dashed-dot (blue), dashed (red) and solid (black) lines, respectively.}
\label{fig6}
\end{figure}

\begin{figure}%
\centering
\includegraphics[scale=0.2]{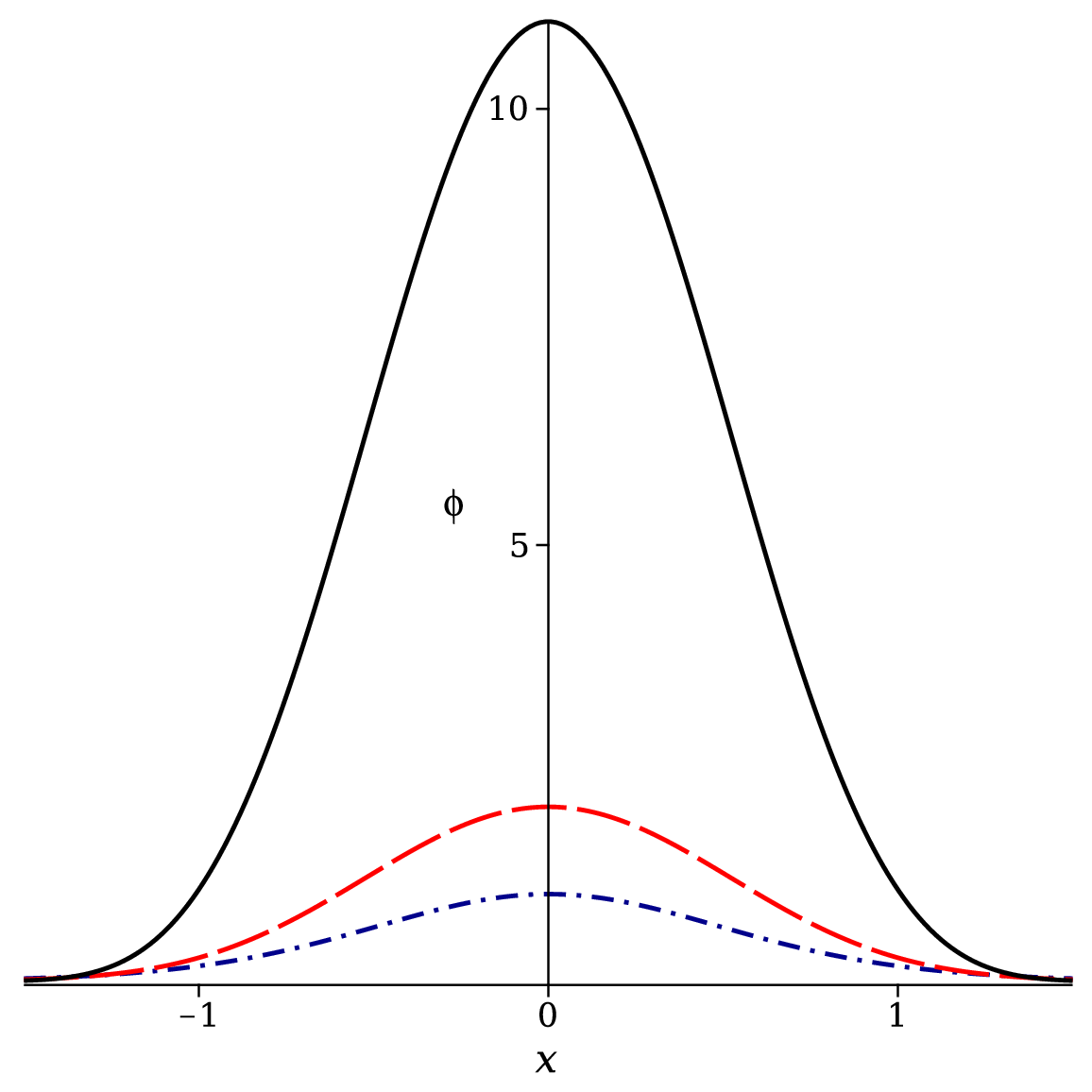}
\includegraphics[scale=0.2]{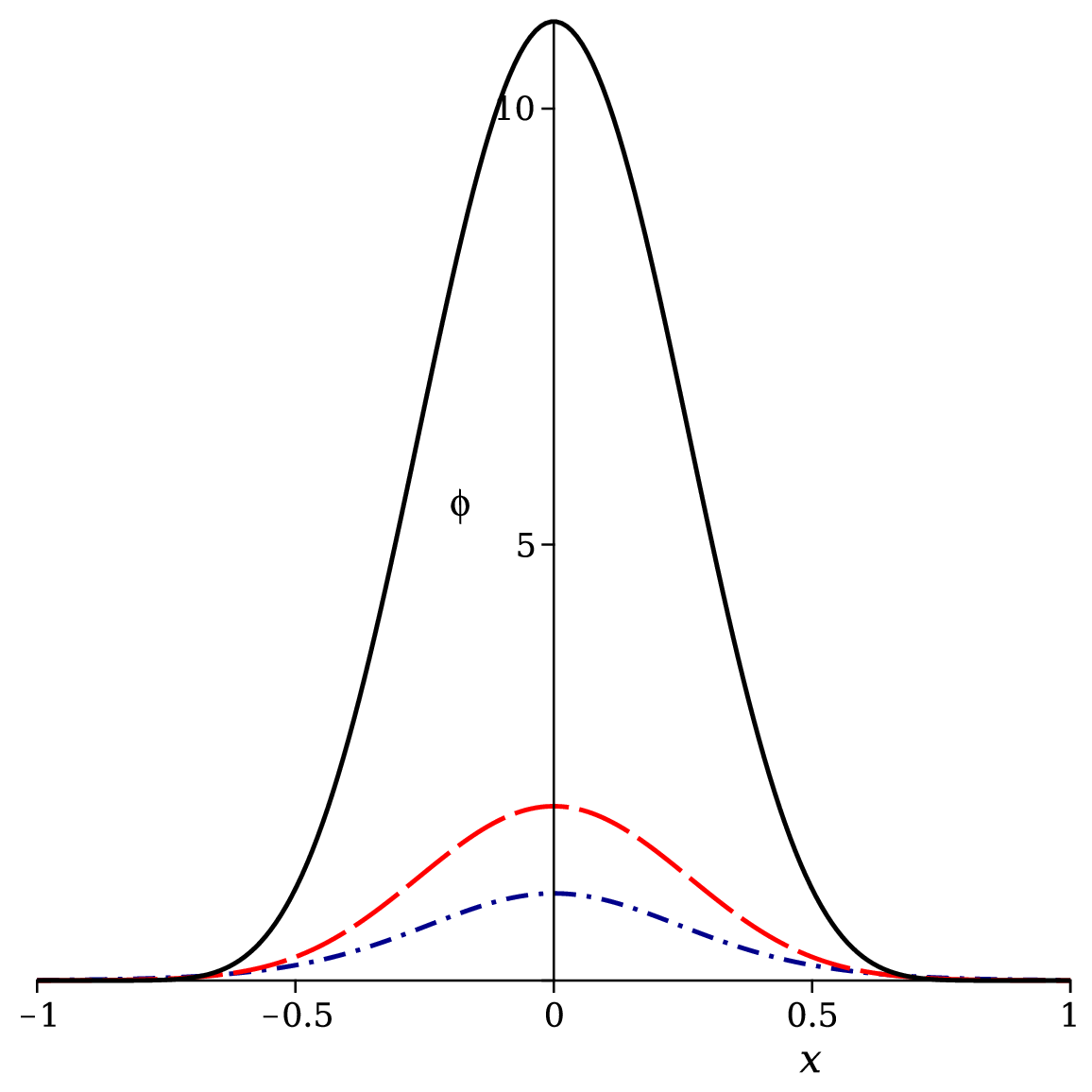}
\caption{Energy densities of the solutions for the general models \eqref{v1mod}  (left panel) and \eqref{Vmod2} (right panel), considering the same values of parameters used in Figures~\ref{fig4} and \ref{fig6}.}
\label{fig8a}
\end{figure}

Therefore, the general model \eqref{Vmod2} provides solutions that smoothly lead \eqref{solstarstandard} into \eqref{solstaromod} as $\alpha$ increases. For $\alpha\rightarrow 0$ we get the kink solution \eqref{solstarstandard} and for $\alpha\rightarrow \infty$ we get the compacton solution \eqref{solstaromod}. This transformation is represented in Fig.~\ref{fig6} for  $n=1$ and some values of $\alpha$, where the solution is going from \eqref{sol1staro} to \eqref{solstarcompacton}. The energy densities of these solutions are depicted on the second panel of Fig.~\ref{fig8a}.

\begin{figure}%
\centering
\includegraphics[scale=0.3]{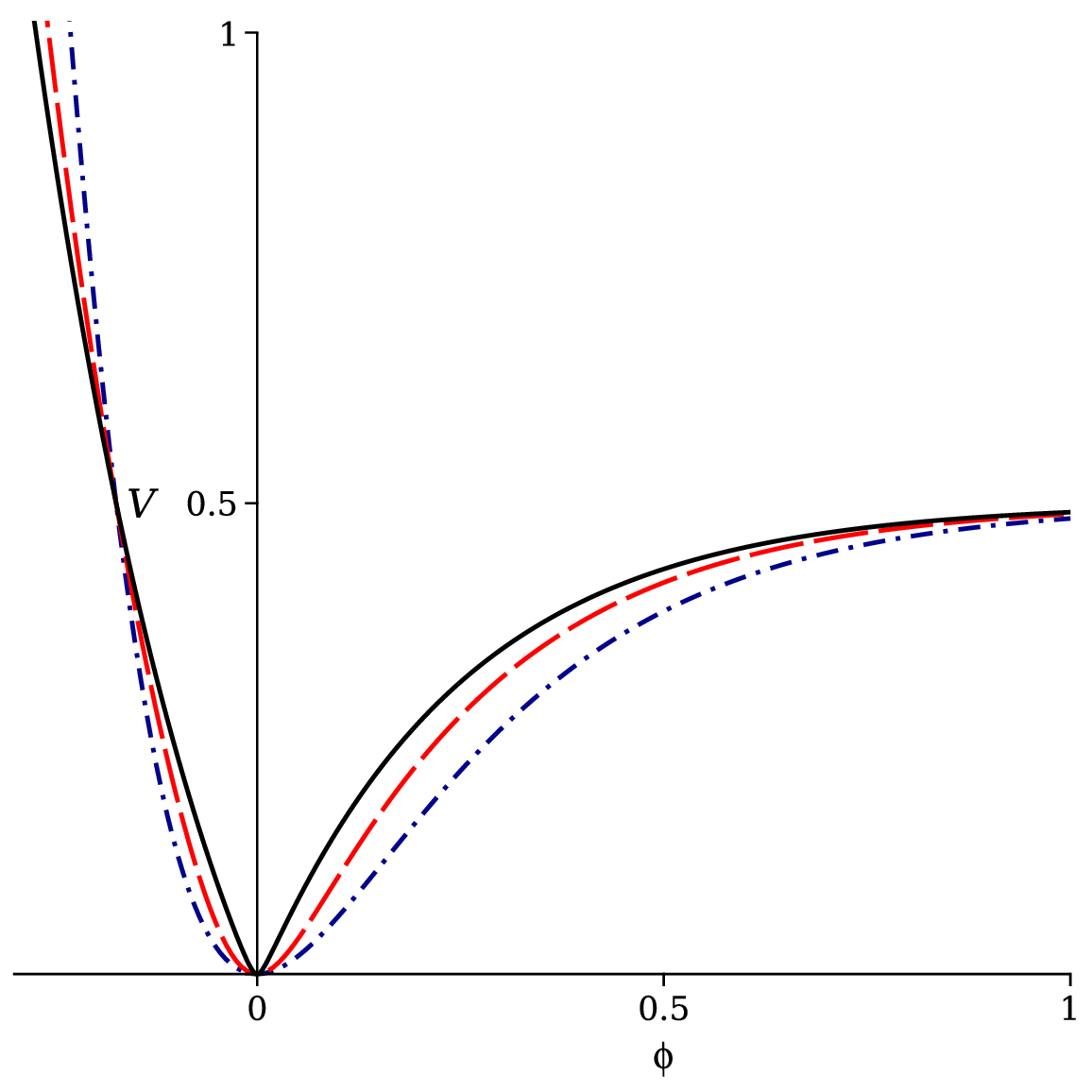}
\caption{Potential $V_{mod}^{Star}(\phi)$ given in \eqref{VmodStar} for $\alpha=0.1,1,10$,  represented by dashed-dot (blue), dashed (red) and solid (black)  lines, respectively.}
\label{fig8}
\end{figure}

 In the case  where $n$ is significantly greater than one, we find the following modification on the original Starobinsky potential  
\be
V_{mod}^{Star}(\phi)=\frac{1}{6\alpha}\left(\sqrt{1+6\alpha \left(1+\frac32\alpha\right)\left(1- \e^{-4\phi}\right)^2}-1\right).
\label{VmodStar}
\ee
Fig.~\ref{fig8} represents the behaviour of the potential above as $\alpha$ increases. At the limit $\alpha\rightarrow 0$, it becomes \eqref{Starobinsky}.

\section{Comments and conclusions}\label{sec-4}
In this work, we considered two interesting potentials to explore kinks and compactons in models governed by a single real scalar field in two-dimensional spacetime. We investigated scenarios where the scalar field exhibits standard kinematics, and then we assumed two possibilities of generalizations of the kinetic term. The first one presents quadratic dynamics, and the second is controlled by a real parameter $\alpha$. The value of $\alpha$ determines whether the generalized model is closer to (for small values) or farther away (for larger values) from the standard model. In this sense, the first two analyzed scenarios (with dynamics $F=X$ and $F=-X^2$) are expressed by a global description, where $\alpha$ controls the evolution from standard dynamics to a regime characterized by compact solutions. This transition occurs gradually, leading to significant changes in the structural properties of the solutions as $\alpha$ increases. For the regime of very large $\alpha$, the modified potentials $V_{mod}^{(1)}(\phi)$ and $V_{mod}^{(2)}(\phi)$ presented a new behaviour as illustrated in Figs.~\ref{fig5} and \ref{fig7}.  Despite this, they preserve the compact solutions of the theory with $F=-X^2$, resulting in well localized energy densities.

These results can be useful in studies involving both topological and cosmological contexts, since the extended models can admit a wider set of solutions. From the topological point of view, one may consider higher-dimensional representations such as domain walls or the five-dimensional braneworld scenario \cite{brane1,brane2,brane3}, to study how the brane behaves under the presence of the solutions found in the present work.   From the cosmological perspective, the modifications of the Starobinsky potential proposed in Section~\ref{sec4b} may satisfy the criteria necessary for describing observational data, mainly during the inflationary phase. Since the defect structures are inhomogeneous solution in space, their cosmological counterparts may appear as inhomogeneous time configurations, inducing spontaneous breaking of time translation symmetry, as considered in Ref. \cite{Wil}. This issue may lead to interesting new investigations, as the ones suggested, for instance, in the cosmic time crystal evolutions studied in Refs. \cite{TC1,TC2,TC3}. Moreover, since the original Starobinsky potential stands as one important possibility to describe dynamics of inflation, incorporating it into a broader scenario appears to be an effective approach for evaluating the robustness of the model in describing inflationary cosmology. These are intriguing points that need further consideration in future studies.

\section*{Acknowledgments}

The authors would like to thank Conselho Nacional de Desenvolvimento Científico e Tecnológico (Grants No. 303469/2019-6 and No. 402830/2023-7), Para\'\i ba State Research Foundation (Grant 0015/2019) and PRPGI/IFBA for the financial support.


\end{document}